\PassOptionsToPackage{pagebackref,bookmarks}{hyperref}
\documentclass[journal]{vgtc}                     

\onlineid{1058}

\vgtccategory{Research}

\vgtcpapertype{Representations \& Interaction}

\graphicspath{{figures/}{pictures/}{images/}{./}} 

\usepackage{tabu}               
\usepackage{booktabs}                 
\usepackage{lipsum}                    
\usepackage{mathptmx}                 
\usepackage{color}
\usepackage{enumitem}
\usepackage{graphicx}
\usepackage{multirow}              
\usepackage{tabularx}
\usepackage{setspace}
\usepackage[normalem]{ulem}
\usepackage{wrapfig}
\usepackage[export]{adjustbox}
\usepackage{url}
\usepackage[utf8]{inputenc}
\usepackage{amsmath}

\newcommand{\eg}{{\it e.g.,\ }}
\newcommand{\etal}{{\it et al.\ }}
\newcommand{\etc}{{\it etc.}~}
\newcommand{\ie}{{\it i.e.,\ }}

\newcommand{\re}[1]{{\color{black} #1}}

\usepackage{hhline}
\definecolor{tablerowcolor}{rgb}{0.667,0.667,0.667 }
\definecolor{tablerowcolor2}{rgb}{0,0,0}

\definecolor{ckey}{rgb}{0.031, 0.031, 0.682}
\definecolor{cvalue}{rgb}{0.2, 0.439, 0.145}
\definecolor{bluecrayola}{rgb}{0.12,0.46,1.0}

\newcommand{\toole}{SceneLoom}

\definecolor{mygreen}{RGB}{0,176,80}
\definecolor{mygreen2}{RGB}{232, 245, 233}
\definecolor{myorange}{RGB}{197, 80, 31}
\definecolor{myblue}{RGB}{0, 76, 193}

\PassOptionsToPackage{svgnames}{xcolor}
\usepackage{xcolor}
\usepackage{xspace, xpunctuate}
\usepackage{listings}  
\usepackage{enumitem}  
\usepackage{multirow}
\definecolor{keywordcolor}{rgb}{0.56, 0.13, 0.00}
\definecolor{ndkeywordcolor}{rgb}{0.05, 0.46, 0.17}
\definecolor{commentcolor}{rgb}{0.41, 0.64, 0.70}
\definecolor{stringcolor}{rgb}{0.25, 0.44, 0.63}
\lstdefinelanguage{TypeScript}{
  keywords={typeof, new, true, false, catch, function, return, null, catch, switch, var, if, in, while, do, else, case, break, boolean},
  morekeywords={[2]{class, export, throw, implements, import, this}},
  identifierstyle=\color{black},
  sensitive=false,
  comment=[l]{//},
  morecomment=[s]{/*}{*/},
  commentstyle=\color{commentcolor}\ttfamily,
  stringstyle=\color{stringcolor}\ttfamily,
  morestring=[b]',
  morestring=[b]"
}
\usepackage{hyperref}
\newcommand{\refappendixspec}{\ifx\hideappendix\undefined\ref{appendix:spec}\else A\fi}
\newcommand{\refappendixprompt}{\ifx\hideappendix\undefined\ref{appendix:prompt}\else B\fi}
\newcommand{\refappendixtest}{\ifx\hideappendix\undefined\ref{appendix:test}\else C\fi}

\usepackage{xspace}

\usepackage{url}

\usepackage[framemethod=TikZ]{mdframed}
\definecolor{mycolor}{RGB}{139, 143, 190}
\definecolor{bgcolor}{RGB}{242, 243, 249}

\newmdenv[
  backgroundcolor=bgcolor,
  topline=false,
  bottomline=false,
  rightline=false,
  leftline=false,
  skipabove=0.5em,
  skipbelow=0.5em,
  innerleftmargin=0.5em,
  innerrightmargin=0.5em,
  innertopmargin=0.5em,
  innerbottommargin=0.5em,
  singleextra={
    \draw[mycolor, line width=0.5pt] ([yshift=0pt]O) -- ([yshift=0pt]O-|P);
    \draw[mycolor, line width=0.5pt] ([yshift=0pt]P-|O) -- ([yshift=0pt]P);
    \draw[mycolor, line width=0.5pt] ([yshift=1.5pt]O) -- ([yshift=1.5pt]O-|P);
    \draw[mycolor, line width=0.5pt] ([yshift=-1.5pt]P-|O) -- ([yshift=-1.5pt]P);
  },
  secondextra={
    \draw[mycolor, line width=0.5pt] ([yshift=0pt]O) -- ([yshift=0pt]O-|P);
    \draw[mycolor, line width=0.5pt] ([yshift=1.5pt]O) -- ([yshift=1.5pt]O-|P);
  },
  firstextra={
    \draw[mycolor, line width=0.5pt] ([yshift=0pt]P-|O) -- ([yshift=0pt]P);
    \draw[mycolor, line width=0.5pt] ([yshift=-1.5pt]P-|O) -- ([yshift=-1.5pt]P);
  }
]{coloredquotation}

\definecolor{output}{RGB}{222,234,240}
\definecolor{mycoloroutput}{RGB}{177,211,229}
\newmdenv[
  backgroundcolor=output,
  topline=false,
  bottomline=false,
  rightline=false,
  leftline=false,
  skipabove=0.5em,
  skipbelow=0.5em,
  innerleftmargin=0.5em,
  innerrightmargin=0.5em,
  innertopmargin=0.5em,
  innerbottommargin=0.5em,
    singleextra={
    \draw[mycoloroutput, line width=0.5pt] ([yshift=0pt]O) -- ([yshift=0pt]O-|P);
    \draw[mycoloroutput, line width=0.5pt] ([yshift=0pt]P-|O) -- ([yshift=0pt]P);
    \draw[mycoloroutput, line width=0.5pt] ([yshift=1.5pt]O) -- ([yshift=1.5pt]O-|P);
    \draw[mycoloroutput, line width=0.5pt] ([yshift=-1.5pt]P-|O) -- ([yshift=-1.5pt]P);
  },
  secondextra={
    \draw[mycoloroutput, line width=0.5pt] ([yshift=0pt]O) -- ([yshift=0pt]O-|P);
    \draw[mycoloroutput, line width=0.5pt] ([yshift=1.5pt]O) -- ([yshift=1.5pt]O-|P);
  },
  firstextra={
    \draw[mycoloroutput, line width=0.5pt] ([yshift=0pt]P-|O) -- ([yshift=0pt]P);
    \draw[mycoloroutput, line width=0.5pt] ([yshift=-1.5pt]P-|O) -- ([yshift=-1.5pt]P);
  }
]{coloredquotationoutput}

\colorlet{punct}{red!60!black}
\definecolor{delim}{RGB}{20,105,176}
\colorlet{numb}{magenta!60!black}

\lstdefinelanguage{json}{
    basicstyle=\footnotesize\ttfamily,
    numbers=left,
    numberstyle=\color{gray}\footnotesize\ttfamily,
    numbersep=4pt,
    tabsize=2,
    showstringspaces=false,
    breaklines=true,
    xleftmargin=16pt,
    literate=
     *{0}{{{\color{numb}0}}}{1}
      {1}{{{\color{numb}1}}}{1}
      {2}{{{\color{numb}2}}}{1}
      {3}{{{\color{numb}3}}}{1}
      {4}{{{\color{numb}4}}}{1}
      {5}{{{\color{numb}5}}}{1}
      {6}{{{\color{numb}6}}}{1}
      {7}{{{\color{numb}7}}}{1}
      {8}{{{\color{numb}8}}}{1}
      {9}{{{\color{numb}9}}}{1}
      {:}{{{\color{punct}{:}}}}{1}
      {,}{{{\color{punct}{,}}}}{1}
      {\{}{{{\color{delim}{\{}}}}{1}
      {\}}{{{\color{delim}{\}}}}}{1}
      {[}{{{\color{delim}{[}}}}{1}
      {]}{{{\color{delim}{]}}}}{1},
}

\definecolor{myblue}{RGB}{47, 85, 151}

\PassOptionsToPackage{svgnames}{xcolor}
\usepackage{xcolor}
\usepackage{xspace, xpunctuate}
\usepackage{listings}  
\usepackage{enumitem}  
\usepackage{multirow}
\definecolor{keywordcolor}{rgb}{0.56, 0.13, 0.00}
\definecolor{ndkeywordcolor}{rgb}{0.05, 0.46, 0.17}
\definecolor{commentcolor}{rgb}{0.41, 0.64, 0.70}
\definecolor{stringcolor}{rgb}{0.25, 0.44, 0.63}
\lstdefinelanguage{TypeScript}{
  keywords={typeof, new, true, false, catch, function, return, null, catch, switch, var, if, in, while, do, else, case, break, boolean},
  morekeywords={[2]{class, export, throw, implements, import, this}},
  identifierstyle=\color{black},
  sensitive=false,
  comment=[l]{//},
  morecomment=[s]{/*}{*/},
  commentstyle=\color{commentcolor}\ttfamily,
  stringstyle=\color{stringcolor}\ttfamily,
  morestring=[b]',
  morestring=[b]"
}
\lstdefinelanguage{mylang}{
}
\lstdefinestyle{mystyle}{
  basicstyle=\footnotesize\ttfamily,
    numbers=left,
    numberstyle=\color{gray}\footnotesize\ttfamily,
    numbersep=4pt,
  frame=none,
  columns=flexible,
  xleftmargin=10pt,
  aboveskip=-1pt,
  belowskip=0pt,
  language=mylang
}

\title{\toole: Communicating Data with Scene Context}

\author{%
  \authororcid{Lin Gao}{0009-0004-1613-1774},
  \authororcid{Leixian Shen}{0000-0003-1084-4912}, 
  \authororcid{Yuheng Zhao}{0000-0003-1573-8772}, 
  Jiexiang Lan,
  \authororcid{Huamin Qu}{0000-0002-3344-9694},
  \authororcid{Siming Chen}{0000-0002-2690-3588}
}

\authorfooter{
    \item
  	L. Gao, Y. Zhao, J. Lan and S. Chen are with Fudan University. S. Chen is also with Ji Hua Laboratory and Shanghai Key Laboratory of Data Science. S. Chen is the corresponding author.
  	E-mail: lingao23@m.fudan.edu.cn, simingchen@fudan.edu.cn.
   \item 
    L. Shen and H. Qu are with the Hong Kong University of Science and Technology.
}

\abstract{
In data-driven storytelling contexts such as data journalism and data videos, data visualizations are often presented alongside real-world imagery to support narrative context.
However, these visualizations and contextual images typically remain separated, limiting their combined narrative expressiveness and engagement.
Achieving this is challenging due to the need for fine-grained alignment and creative ideation.
To address this, we present SceneLoom, a Vision-Language Model (VLM)-powered system that facilitates the coordination of data visualization with real-world imagery based on narrative intents.
Through a formative study, we investigated the design space of coordination relationships between data visualization and real-world scenes from the perspectives of visual alignment and semantic coherence. 
Guided by the derived design considerations, SceneLoom leverages VLMs to extract visual and semantic features from scene images and data visualization, and perform design mapping through a reasoning process that incorporates spatial organization, shape similarity, layout consistency, and semantic binding.
The system generates a set of contextually expressive, image-driven design alternatives that achieve coherent alignments across visual, semantic, and data dimensions.
Users can explore these alternatives, select preferred mappings, and further refine the design through interactive adjustments and animated transitions to support expressive data communication.
A user study and an example gallery validate SceneLoom's effectiveness in inspiring creative design and facilitating design externalization.
}


\teaser{
  \centering
  \includegraphics[width=\linewidth]{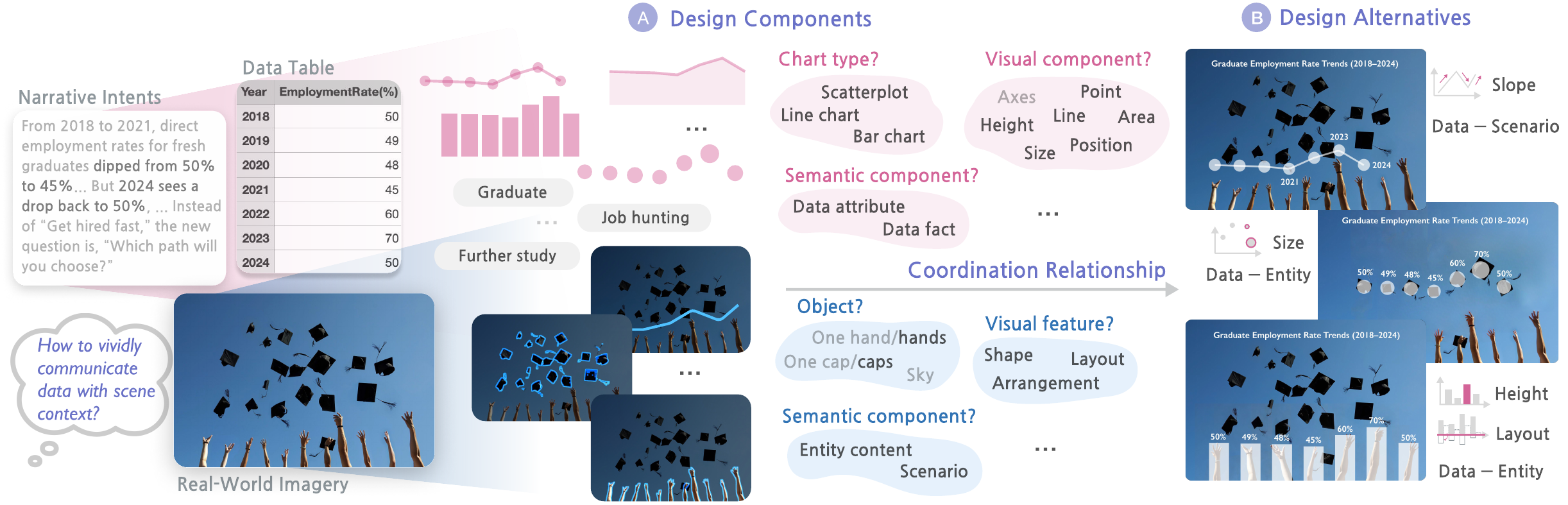}
  \caption{%
  SceneLoom explores creative ways to blend data visualization with real-world scene context for expressive data-driven storytelling. Given narrative intents and data, SceneLoom (A) bridges the complex and diverse design space between data visualization and scene imagery, and (B) generates contextually expressive design alternatives.
  }
  \label{fig: teaser}
}

\begin{document}

\begin{spacing}{0.97}
\firstsection{Introduction}
\maketitle


In data-driven storytelling practices (\eg data journalism or data videos), real-world scenes and data visualizations serve as two foundational visual elements, each contributing in distinct yet complementary ways to the overall narrative~\cite{shen2025reflecting}. 
Real-world scenes, \ie images or footage of environments, events, or activities, can provide spatial and temporal context~\cite{tang2022smartshots}, evoke emotional resonance~\cite{lan2024affective}, and offer visual cues that can inform the design of accompanying visualizations~\cite{kang2021metamap, choi2024creative}. 
Meanwhile, \re{in the data-driven storytelling context, data visualizations often serve to convert abstract information into graphical representations that highlight patterns and insights central to the narrative.}
Although complementary, real-world scenes and data visualizations differ in fundamental ways.
Real-world scenes often convey subjective narratives through concrete imagery, while data visualizations \re{can} encode abstract data relationships and emphasize factual clarity.
As shown in Fig.~\ref{fig: teaser}, the scene depicts the celebratory moment of tossing graduation caps, conveying emotional and subjective intent, while the data visualization presents objective data on graduate employment.
These modalities differ in information type, perceptual mode, and communicative goals, introducing distinct design considerations in areas such as semantics and visual features.
Such differences make their seamless integration particularly challenging.

Recent studies have explored integrating real-world elements into data visualization as foreground objects or background canvases to enhance visual expressiveness~\cite{xiao2024chart}. These approaches either rely on data attribute mapping~\cite{zhang2020dataquilt} or visual feature alignment~\cite{kouts2023lsdvis}, with many leveraging text-to-image models to reinforce semantic consistency~\cite{coelho202infomage, wu2023viz2viz}. However, these methods often ignore the systematic understanding of image content and data visualization, thus lacking fine-grained alignment. For instance, they tend to overlook critical structural components within visualizations, such as coordinate systems and spatial layouts, as well as the rich narrative contexts inherent in real-world scenes, like spatial relationships and character roles. 
Without a structured framework to analyze and align the expressive dimensions of both modalities, such approaches limit the space of design possibilities and constrain the system’s ability to support open-ended exploration and creative design.

Therefore, coordinating real-world scenes with data visualizations remains a significant challenge.
As noted earlier, the two modalities differ in the types of elements they contain and the design dimensions for effective coordination. Such divergence makes it inherently difficult to construct a systematic understanding of their components and the potential mappings between them, thereby limiting the space for creative design.
Moreover, narrative-driven coordination goes beyond simply overlaying or juxtaposing visuals. The key challenge lies in simultaneously accounting for shared aspects (\eg spatial structures, thematic focus) and resolving inconsistencies (\eg mismatched element correspondences and conflicting narrative cues). Achieving this level of coordination requires a deep understanding of elements, their semantic roles, and spatial positioning, which is particularly challenging in the context of complex real-world scenes.
Meanwhile, although Vision-Language Models (VLMs) excel at general visual understanding, they struggle to grasp the underlying design logic and contextual reasoning needed for effective coordination without additional knowledge.


To address these issues, we first conducted a formative study to analyze design components in data visualizations and real-world scenes and derive a set of coordination relationships from visual and semantic perspectives.
Building on these insights, we developed SceneLoom, a prototype system that enables context-aware coordination between data visualizations and real-world scenes based on narrative intents, resulting in expressive and creative design outcomes.
Given narrative text, a data table, and real-world imagery, SceneLoom begins with data preparation to extract relevant design components. 
The VLM-powered coordination process is structured in two stages: perception and reasoning. 
In the perception stage, components are specified along key dimensions of the design space to support VLM interpretation.
In the reasoning stage, the VLM performs design mapping guided by a set of derived design considerations, including spatial organization, shape similarity, layout consistency, and semantic binding.
To support fine-grained alignment, SceneLoom enables visualization adjustment and image editing to resolve data-element conflicts. 
After user refinement, the system further generates animated transitions for visualization elements to enhance narrative flow. 
We evaluated the system through a user study and curated an example gallery to demonstrate its expressive potential.
Our main contributions are as follows:
\begin{itemize}[nosep,leftmargin=*]

\item A design space that identifies design components and coordination relationships between real-world scenes and data visualizations to support visual alignment and semantic coherence.

\item SceneLoom, a prototype system that supports context-aware coordination between real-world scenes and data visualizations. It integrates VLM-powered perception and reasoning to enable fine-grained alignment and creative support.

\item An example gallery and a user study to validate the expressiveness and effectiveness of SceneLoom.
\end{itemize}

\section{Related Work}
This section reviews related work on blending data with real-world elements, image-driven creative tools, and VLM-based visual reasoning.

\subsection{Blending Data Visualization with Real-World Elements}



With growing attention to the physical world~\cite{herman2025touch}, as well as personal~\cite{huang2015personal} and societal data~\cite{morais2022showing}, embedded visualization has expanded beyond Augmented Reality (AR)~\cite{willett2017embedded} into 2D contexts, enhancing links between real-world elements and data.

In 2D settings, recent studies have also explored ways to embed real-world elements or scenes into visual representations. For example, Infomage~\cite{coelho202infomage} integrates data visualizations into thematic images through image processing and visual distortion optimization.
DataQuilt~\cite{zhang2020dataquilt} extracts visual elements from raster images and binds them to data via an iterative process.
Beyond Numbers~\cite{chen2024beyond} aligns scene elements with data through visual analogy, but departs from traditional charts in favor of naturalistic representations.
With the rise of generative models~\cite{zhang2023controlnet, rombach2022stable, avrahami2022inpainting}, text-to-image techniques have further eased the integration of real-world elements. Several tools~\cite{wu2023viz2viz, kouts2023lsdvis, xiao2024chart} extract visualization features and incorporate them as backgrounds or foreground inputs for generative models, using deep optimization and conditional generation to reduce uncertainty and improve control.

Despite progress, prior work mainly emphasizes visual presentation, often neglecting the narrative role of real-world scenes.
\re{While domain-specific applications, such as those in sports analytics~\cite{yao2024motion, zhutian2022viscommentator}, enhance context understanding, they are limited by specific design spaces~\cite{yao2025user, zhutian2023sporthesia}.}
The absence of a general design space hinders the diversity and expressiveness of integrated outcomes. \re{We bridge this gap through a systematic analysis of modality components, identifying their core design elements and coordination relationships.}

\subsection{Creative Support Tools by Image-Driven Inspiration}




Images are the multifaceted source of inspiration for designers and developers throughout ideation~\cite{simon2016ideation, choi2024creative}, exploration~\cite{ye2024visatlas}, and prototyping~\cite{cheng2023prior}. Building on this, image-driven creativity-supporting tools are widely applied in graphic design, digital art, and storytelling.

Recent studies on reference images have primarily focused on visual and semantic aspects.
Visual features, such as color palettes~\cite{shi2024mondrian, yuan2022infocolorizer}, textures~\cite{lukac2013painting}, and styles~\cite{zhou2024stylefactory}, are mapped more abstractly at a global level, shaping the overall perception of an image while conveying emotion. 
Geometric structures are mapped based on similarity and coherence. Chen~\etal~\cite{zhutian2020timeline} identified compositional patterns in timeline infographics to inform new designs. Chilton~\etal ~\cite{chilton2019visiblends, chilton2021visifit} applied shape constraints to enable visual blending.
Semantics serve not only as prompts for retrieval and generation but also as conceptual anchors, capturing entities and contexts that guide visual reinterpretation~\cite{choi2024creative}.
Moreover, combining multiple images introduces new dimensions of creativity. Like MetaMap~\cite{kang2021metamap}, the relative positioning, shared features, and mapping of distances between images can spark unexpected ideas.

\re{Regarding the mapping methods, Brickify~\cite{shi2025brickify} requires abstraction for free-form generation, and Data Pictorial~\cite{zhou2024pictorial} requires extracting precise element data to further input into generative models for next-step generation.} With the advancement of LLMs, image-driven creativity has evolved to emphasize iterative refinement~\cite{zhou2024understanding}, alongside traditional stages such as brainstorming and alternatsive filtering.

While prior work highlights the creative potential of image-driven tools, most focus on open-ended mappings for exploratory or artistic use, with minimal design constraints. As a result, they often lack structured reasoning frameworks suited for goal-oriented, constrained scenarios such as data visualization.
In our work, we focus on visualization contexts and examine which visual and semantic features in reference images can effectively inform design. By analyzing real-world elements across varying granularities and dimensions, we position images as sources of inspiration and carriers of narrative meaning.


\subsection{VLM-driven Visual Understanding and Reasoning}



VLMs extend Large Language Models (LLMs) with visual encoding, enabling models to ``see'' and perform tasks such as image captioning~\cite{Guo2024regiongpt}, visual question answering~\cite{ChartInsights}, and visual reasoning~\cite{liu2023visual}. To enhance reasoning capabilities, recent works integrate traditional image processing~\cite{kuruvilla2016review} or deep learning methods~\cite{kirillov2023sam, li2024semanticsam} to provide language-guided image tokens~\cite{wang2023visonllm}, supporting downstream tasks such as planning and tool execution~\cite{yang2023mmreact, wang2024genartist, li2024situationadapt}.

As a special form of visual representation, data visualization presents unique challenges for VLMs~\cite{zeng2025advancing}. Lundgard~\etal~\cite{lundgard2022accessible} identified four levels of semantic understanding for data visualization: visual elements and properties, statistical concepts and relationships, graphical perception, and contextual or domain-specific insights. \re{Recent work~\cite{islam2024lvlms} has explored the ability of VLMs to understand charts across these levels in various downstream tasks.}
ChartInsighter~\cite{wang2025chartinsighter} investigates VLMs' understanding of time series charts, focusing on the first two levels. Guo~\etal~\cite{guo2024understanding} examine graphical perception tasks, such as position, height, and angle. Tasks requiring contextual understanding are often studied in domain-specific scenarios~\cite{liu2025smartboard}, while research on general storytelling focuses more on semantic coherence~\cite{li2025composing}.

To support contextual understanding, we translate design space dimensions into structured specifications, enabling VLMs to better interpret task-relevant visual elements. 
While prior work, such as Meng~\etal~\cite{meng2025mmiu}, has explored multi-image understanding in natural scenes, limited attention has been given to cross-modal relationships, particularly between data visualizations and natural images.
To address this gap, we introduce a coordination process comprising perception and reasoning to guide VLMs in forming meaningful connections.

\section{Formative Study}

\begin{figure*}[htbp]
  \centering 
  \includegraphics[width=\linewidth]{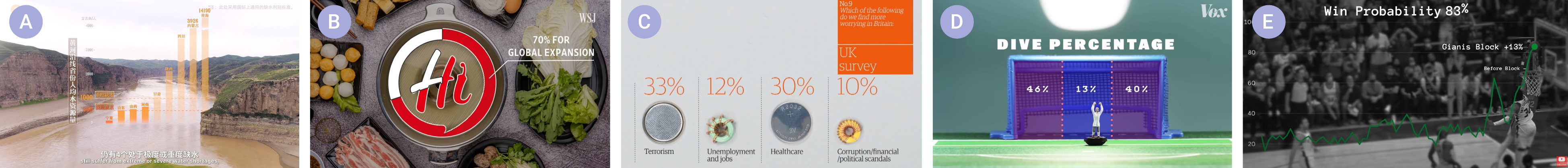}
  \caption{Examples of integration cases. (A)Residential water resources along the Yellow River, using the riverbed as a baseline~\cite{xinpianchang2024yellowriver}. (B)Fund allocation with a pie chart matching the hot pot shape~\cite{wsj2024haidilao}. (C)Public concerns in the UK, represented by the size of physical objects~\cite{guardian2014uk}. (D)Goalkeeper dive percentages mapped to the goal layout~\cite{vox2023soccer}. (E)Win probability changes aligned with a key basketball dunk~\cite{mac2024basketball}.
  }
  \label{fig: corpus}
  \vspace{-0.4cm}
\end{figure*}

This section introduces the research question and our formative study to investigate it, as well as the derived design space and coordination strategies, which were further verified by design experts.

\subsection{Research Question} \label{sec: RQ}
Data visualizations and real-world scenes differ fundamentally in information type, perception modes, and communicative goals. This divergence creates tensions in their coordination: (1) semantic gaps between abstract data encoding and concrete scene semantics, and (2) perceptual competition when visual channels overlap.
Therefore, these challenges lead to our central research question: \textit{How to coordinate design components from data visualization and real-world imagery in a narrative-driven context?} 
\re{To explore this, we focus on video-based storytelling, where temporal continuity and frame-by-frame structure reveal how visual elements evolve and interact with real-world scenes over time. Sequencing and camera motion enable the gradual introduction of elements, exposing transitions and coordination patterns.}
\re{Building on this temporal structure, we analyze existing video cases to examine how data visualizations interact with real-world scenes (Sec.~\ref{sec: corpus analysis}). In Sec.~\ref{sec: design components}, we conducted frame-by-frame analysis to identify design components from both domains. Based on these findings, Sec.~\ref{sec: coordination} presents coordination strategies that govern how these components are visually and semantically linked.}

\subsection{Corpus Analysis}\label{sec: corpus analysis}
\re{To ground our research question in concrete examples and derive actionable insights for subsequent analysis, we conducted a corpus analysis of existing data videos. We first surveyed data videos from prior studies~\cite{yang2022designspace, shi2021comunicating}, reputable news agencies (\eg Vox, BBC News), and major platforms (\eg YouTube, TikTok). Using keywords such as ``data-driven stories'' and ``real-world data videos'', we initially selected cases that featured tight and creative integration between data visualizations and real-world scenes. Due to limited cases, we broadened our scope to include videos on real-world topics with co-occurring visualizations, even if loosely coupled. In total, we selected 54 videos that featured both data visualizations and real-world elements.}

\re{These cases helped us extract common patterns of interaction and define a set of analytical dimensions. Guided by prior work~\cite{willett2017embedded, tong2024vistellvr}, we iteratively refined the coding scheme. Adopting an abductive coding approach, two authors independently coded the videos and resolved differences through discussion, resulting in six analytical dimensions.
\textit{Visualization components} capture how real-world elements are involved in the data graphics, categorized into coordinate systems (15), marks (26), and annotations (16). \textit{Image components} describe real-world elements shown in the scene; due to their diversity, these were annotated in free-text form. \textit{Compositional layout} captures whether the visualization appeared in the foreground (40) or background (14). \textit{Visual inspiration} reflects perceptual connections such as shared shape (11), size (13), or position (11). \textit{Semantic inspiration} refers to conceptual or thematic links, drawn from either metadata (31) or data context (23). \textit{Narrative intent} refers to the communicative goal of the visualization-scene pairing, including explanation (26), comparison (20), and emphasis (9). 
These cases and coding results are available online\footnote{\url{https://airtable.com/apparxcuOrUlTeKj3/shrFcnYr0QytfWLeE}}. Representative examples are shown in Fig.~\ref{fig: corpus}.}

\subsection{Identifying Design Components} \label{sec: design components}

\begin{figure}[h]
  \centering 
  \includegraphics[width=\columnwidth]{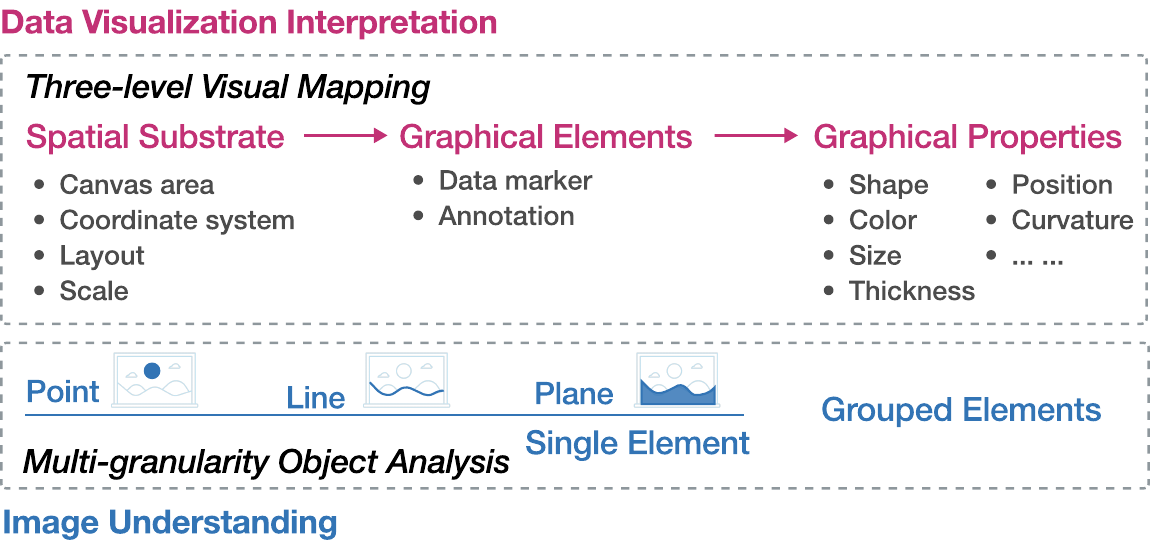}
  \caption{Visual components in data visualization and real-world images.}
  \label{fig: design elements}
  \vspace{-0.4cm}
\end{figure}

\re{Based on insights from the corpus analysis and related literature~\cite{zhutian2022viscommentator, shi2025brickify}, we analyze design components from two key aspects: visual cues and semantic content. These aspects often intertwine in practice, as shown in Fig.\ref{fig: corpus}A–E. Guided by this observation, we examine components from both the visualization side (Sec.\ref{sec: vis components}) and the real-world scene side (Sec.\ref{sec: real-world components}). The resulting visual components are summarized in Fig.\ref{fig: design elements}.}

\subsubsection{Design Components from Data Visualization} \label{sec: vis components}
In analyzing data visualization components, we identify visual components through \re{a visual mapping framework~\cite{card1999readings}} and characterize the semantic content conveyed by the data.

\textbf{Visual Components.}
Previous work~\cite{ying2022glyphcreator, zhutian2020timeline, shen2023towards} has analyzed data visualizations from multiple perspectives, including form-related features.
\re{We build on the influential framework by Card~\etal~\cite{card1999readings}, which is rooted in visual perception and breaks down visualizations into three components: \textit{Spatial Substrate}, \textit{Graphical Elements}, and \textit{Graphical Properties}. This framework forms the analytical foundation. Building on corpus insights, we extend it through a fine-grained analysis of components that support alignment with real-world scenes.}
\begin{itemize}[parsep=0.1pt]
    \item \uline{\textit{Spatial Substrate.}} This refers to the foundational space where data visualizations are constructed.  It encompasses data dimensions, axes, graphical boundaries, and spatial layout configurations. We categorize these as \textit{canvas area}, \textit{coordinate system}, \textit{layout} and \textit{scale}.
    \item \uline{\textit{Graphical Elements.}} These include geometric representations such as points, lines, and areas. In addition to \textit{data markers}, they also include \textit{annotations} such as gridlines and shaded regions.
    \item \uline{\textit{Graphical Properties.}} These mainly involve encoding channels, including \textit{size}, \textit{color}, and \textit{shape}, to represent data dimensions or categories, or to emphasize specific aspects of the data. Different mark types involve distinct encoding properties. For example, line marks can be characterized by attributes such as \textit{slope}.
\end{itemize}

\textbf{Semantic Components.}
Data visualization conveys information and insights from the underlying data and its descriptive context. We refer to this as \textit{Data Content}.
\begin{itemize}[parsep=0.1pt]
\item \uline{\textit{Data Content.}} This includes data attributes, data values, and contextual information. As data visualizations are often shaped by user intent and narrative goals, we also consider data facts as part of the content.
\end{itemize}

\subsubsection{Design Components from Real-world Scene} \label{sec: real-world components}
While computer vision has made progress in recognizing structural and semantic elements in real-world scenes, a unified classification for systematic analysis remains absent.  Drawing on case studies and observations of complex scenes, we analyze visual composition and interpret meaning at multiple levels of granularity.

\textbf{Visual Components.}
The gathered examples indicate that design components in real-world scenes exhibit varying levels of granularity, ranging from the entire scene as a background (Fig.~\ref{fig: corpus}A, E) to groups of elements representing data series (Fig.~\ref{fig: corpus}C), or individual elements (Fig.~\ref{fig: corpus}B, D).  
Accordingly, we adopt a multi-granularity perspective and categorize real-world design elements into two types: the \textit{Single Element} and \textit{Grouped Elements}.

\begin{itemize}[parsep=0.1pt]
    \item \uline{\textit{Single Element.}} This refers to an individual element, which can be a \textit{point}, \textit{line}, or \textit{plane}. A point can represent a single physical object with no size constraints, but it does not represent a region, as the basketball in Fig.~\ref{fig: corpus}E. A line can be a physical line or a visually implied line formed by the arrangement of elements. In Fig.~\ref{fig: corpus}A, the visual line formed by the river surface is considered. A plane represents a defined region within the scene or can serve as a mask for an object, defining its boundaries or spatial extent. The examples in Fig.~\ref{fig: corpus}B and D all serve as the planes.
    \item \uline{\textit{Grouped Elements.}} These are combinations of multiple single elements organized with some logical coherence. Due to the diversity and complexity of grouping methods, our work primarily focuses on combinations where the shapes share similarities and have strong semantic associations. For instance, in Fig.~\ref{fig: corpus}C, the elements with similar circle shapes are considered together.
\end{itemize}

\textbf{Semantic Components.} For the semantic interpretation of real-world scenes, we should focus on the scene as a whole to convey the \re{data context} and the description of specific entities within the scene. So, we approach the semantic analysis from two angles: 
\begin{itemize}[parsep=0.1pt]
    \item \uline{\textit{Entity Objects.}} We describe their physical or logical meanings within the narrative context, like the ``hot pot'', ``coil'', ``basketball'' and ``goal'' in Fig.~\ref{fig: corpus}.
    \item \uline{\textit{Scenario.}} We focus on elements such as time, location, events, and environment, which determine the interpretation of data and convey the background context. In Fig.~\ref{fig: corpus}A, the landscape of the Yellow River indicates the background of the water resource data.
\end{itemize}

\subsection{Coordinating Design Components}\label{sec: coordination}
\re{Building on the identified design components, we outlined coordination strategies along two dimensions (visual alignment and semantic coherence) to guide the integration of data and scene in narrative context.}

\begin{figure}[h]
\vspace{-0.1cm}
  \centering 
  \includegraphics[width=\columnwidth]{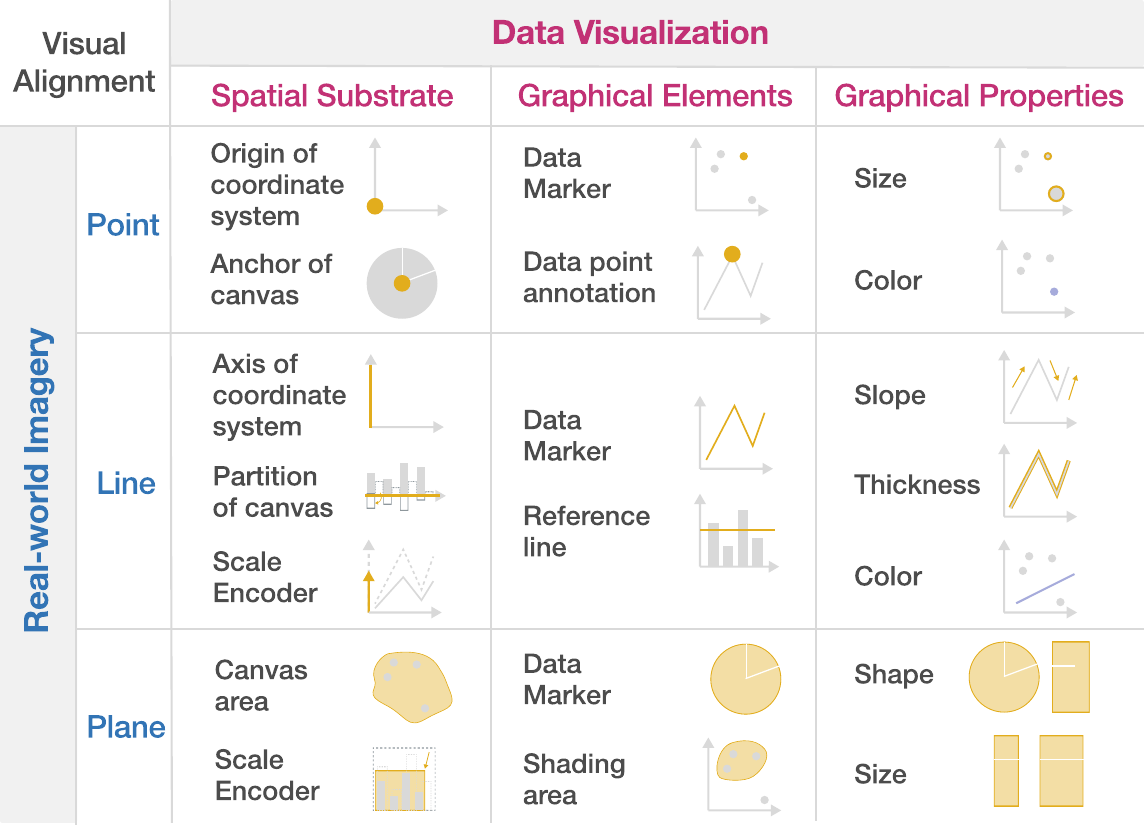}
  \caption{Design space for visual alignment between data visualization and real-world imagery.}
  \label{fig: visual alignment}
  \vspace{-0.3cm}
\end{figure}

\textbf{Visual Alignment.} Building on the decomposition of visual components, we systematically mapped elements from real-world scenes to the data visualization space by considering spatial layout and intrinsic attributes (Fig.~\ref{fig: visual alignment}). In this three-level visual mapping process, real-world design components either directly serve as visualization elements or inform their generation based on contextual roles.

\uline{\textit{Point.}} In the \textit{spatial substrate}, points typically serve as positional anchors, such as coordinate origins or canvas reference points to support spatial alignment.
Within \textit{graphical elements}, shape similarity enables points to map naturally onto point-based data markers. These may also serve as dot-like annotations in line charts or scatterplots to emphasize specific values. In Fig.\ref{fig: corpus}E, a basketball aligns with a highlighted data point through layout matching. Highlighting further utilizes the \textit{properties} of points, such as color and size, as encoding channels. In Fig.\ref{fig: corpus}C, pole size variation visually encodes data magnitude.

\uline{\textit{Line.}} Lines are commonly used to represent connections, outlines, trends, and directional flows. In the \textit{spatial substrate}, they may align with coordinate axes, serving as structural baselines, as illustrated in Fig.~\ref{fig: corpus}A. When layout consistency is maintained, lines naturally divide the canvas, requiring visualization designs to consider symmetry and potential transformations. Line length can also encode scale, influencing proportion and orientation.
As \textit{graphical elements}, lines appear as data markers, such as trend lines or reference lines that highlight baselines, thresholds, or comparisons. Their \textit{graphical properties}, including slope, thickness, and color, help convey relationships and hierarchies. For example, overhead views of highways with varying widths can inspire area charts for traffic volume in network visualizations.

\uline{\textit{Plane.}} Planes support spatial alignment by defining canvas regions and shaping the boundaries within data. As \textit{graphical elements}, they can serve as data markers illustrating distributions or shaded areas that categorize data subsets, as shown in Fig.~\ref{fig: corpus}B and D.
Their \textit{graphical properties}, such as shape and size, influence the structure of area charts and encode scale and magnitude. For example, the shape of the pot in Fig.~\ref{fig: corpus}B informs the design choice for the shape of the area chart.

\re{In practice, visual alignment is primarily achieved through shape similarity and spatial correspondence between real-world elements and visualization components. The most direct way to support such alignment is through overlay or substitution, which allows data elements to be precisely positioned within the scene. These methods were also the most commonly observed in our corpus analysis and form the basis for deriving generalizable coordination principles.}

\textbf{Semantic Coherence.} 
To achieve semantic continuity, the integration of data visualizations and real-world scenes can be approached through two types of mappings:

\uline{\textit{Data Content -- Entity Objects.}} Our case analysis reveals that semantic associations occur at varying levels. We categorize them into three types: \textit{directly indicating data meaning}, \textit{metaphorically representing data}, and \textit{providing contextual information}. 
\re{While the degree of association differs, all levels help bind data to objects, facilitating the instantiation of data concepts and enhancing the expression of data attributes.}
\re{For example, in Fig.~\ref{fig: corpus}D, the goal area in the soccer field directly corresponds to numerical data (\eg goal distribution), serving as a visual representation of specific data columns. In contrast, Fig.~\ref{fig: corpus}A uses the shape of a river to guide the flow of a bar chart, offering contextual cues rather than directly encoding specific data points.}

\uline{\textit{Data Content -- Scenario.}} Contextual information in the scene complements data by situating it within real-world scenarios, emphasizing temporal, spatial, and contextual factors. Fig.~\ref{fig: corpus}E effectively leverages the moment of a slam dunk to complete the data narrative.

\subsection{Expert Interview and Feedback}\label{sec: expert interview}
To validate the design space, we conducted 40-minute \re{semi-structured} interviews with two experienced information visualization experts. One (E1) has over seven years of experience in data journalism, and the other (E2) has five years of experience focused on creative visual communication. Both \re{experts are external to the author team and} offered insightful and constructive feedback.
First, they strongly affirmed the significance of the research problem, noting that creatively aligning data visualizations with scene context is both meaningful and challenging. E1 highlighted that in her TV reporting work, \textit{``it is often necessary to enhance content with data while preserving the authenticity of the scene,''} which fosters audience trust and emotional engagement.

Both experts noted that placing data visualizations in the foreground over real-world scenes is a common and effective design strategy. After reviewing our proposed design space, they agreed it offers a clear framework for linking scene elements with visual representations and found the identified combinations both valid and practical. E1 emphasized the importance of creativity, stating, \textit{``In practice, many design ideas are inspired not only by the content itself but also by existing examples.'' }This highlights the value of presenting design alternatives through recommended templates to inspire users and support effective design decisions.
Meanwhile, both experts are concerned about uncertainties in the design process, especially when scene images lack clear visual or semantic cues. \textit{``In extreme cases''}, E1 noted, \textit{``fallback strategies are needed.''} E2 further stressed the need for interactive operations, \textit{``Designers should be able to adjust and refine recommended results to improve readability and better meet their goals.''} To support flexibility, we incorporate user manipulation features into our prototype system, enabling freeform creation and interactive refinement.

Based on this feedback, we further recognized the role of the design space in guiding design generation. In Sec.~\ref{sec: visual perception}, we detail how data charts or images can be specified into actionable design representations and present key design considerations. 
\begin{figure*}[htbp]
  \centering 
  \includegraphics[width=\linewidth]{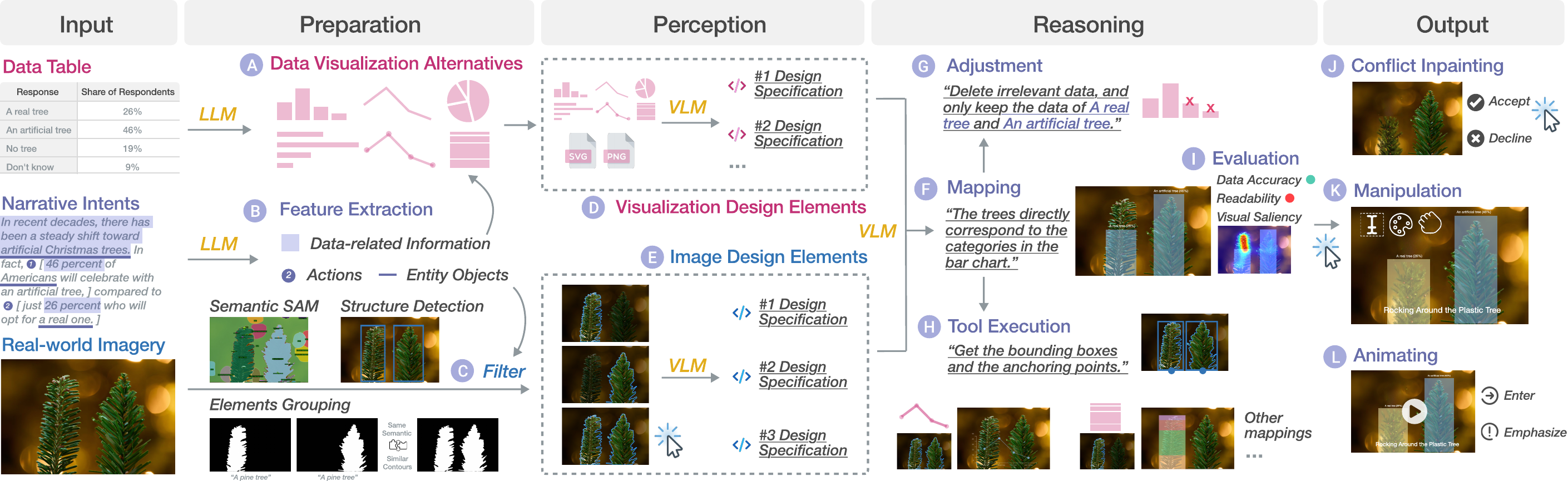}
  \caption{The SceneLoom workflow for coordinating real-world imagery and data visualization based on narrative intent. It consists of five stages: Input, Preparation, Perception, Reasoning, and Output. A Christmas tree preference survey is used as an example to demonstrate the process.}
  \label{fig: workflow}
  \vspace{-0.4cm}
\end{figure*}

\section{SceneLoom}\label{sec: methods}

In this section, we first give an overview of SceneLoom and then go through the workflow with an example in detail (Fig.~\ref{fig: workflow}). The interface and the outcomes of SceneLoom are illustrated in Sec.~\ref{sec: interface}.

\subsection{Workflow}

To achieve visual alignment and semantic coherence during coordination, we propose a VLM-assisted workflow (Fig.~\ref{fig: workflow}) comprising data preparation, visual perception, and reasoning. 
\re{The workflow takes structured data (CSV), narrative text, and real-world images (PNG/JPG) as input.}
In data preparation, \re{SceneLoom} extracts key narrative features (Fig.~\ref{fig: workflow}A), generate visualizations (Fig.~\ref{fig: workflow}B), and filter relevant images based on content and layout structure (Fig.~\ref{fig: workflow}C).
\re{Filtered elements are treated as design components, and VLMs extract their visual attributes using a standardized specification format (Fig.~\ref{fig: workflow}D-E).}
VLMs follow design considerations to guide the mapping process (Fig.~\ref{fig: workflow}F). To achieve fine-grained alignment, visualization adjustment is incorporated during reasoning (Fig.~\ref{fig: workflow}G), \re{and final mappings are executed through LLM-driven tool invocation (Fig.~\ref{fig: workflow}H).}
Design alternatives are \re{automatically evaluated by VLMs} for data accuracy and visual communication effectiveness (Fig.~\ref{fig: workflow}I), enabling user selection. 
\re{The interface also supports optional image editing to address inconsistencies (Fig.~\ref{fig: workflow}J).} Once the design is selected, users can interactively refine the canvas (Fig.~\ref{fig: workflow}K), and \re{SceneLoom continuously generates} aligned animations (Fig.~\ref{fig: workflow}L).

Our workflow integrates state-of-the-art models to ensure the accuracy and robustness of outputs. \re{Segment Anything Model (SAM)~\cite{kirillov2023sam}, Semantic-SAM~\cite{li2024semanticsam}, Holistically-Nested Edge Detection (HED)~\cite{ren2012hed}, and Mobile-Lite Structure Detector (M-LSD)~\cite{gu2022mlsd} are used for image processing, and OpenAI's GPT-4o~\cite{achiam2023gpt} supports visual understanding, reasoning, and code interpretation.} 
\re{Interactions with LLMs are implemented via natural language prompts that specify analysis goals, design constraints and generation tasks. Sample prompts and implementation details are provided in Appendix, which is included in the supplementary materials. }

\re{In following sections, we illustrate each stage using a case study from a 2024 U.S. survey on Christmas tree preferences\footnote{\url{https://www.nationalgeographic.com/environment/article/history-origin-artificial-Christmas-trees}}. The data highlights shifting consumer choices, including real trees, artificial trees, or no purchase. The narrative particularly emphasizes a growing preference for artificial trees over real ones.}

\subsection{Data Preparation}\label{sec: data preparation}
Given the complexity of image content and the diversity of data visualization, data preparation is essential. We adopt a narrative-driven approach in which the user's intent is first interpreted to uncover design-relevant cues. These cues then inform the generation of data visualizations and the selection of image elements.

\textbf{Feature Extraction.} Narrative intents play a central role \re{throughout the process}. As shown in Fig.~\ref{fig: workflow}B, \re{SceneLoom extracts features such as data-related content, actions, and entity objects from the data table and input narration}. \re{In addition to identifying values and attributes, the system captures data facts to inform appropriate visualization mappings (\eg trends, comparisons, or distributions)}. Actions, particularly enter and emphasis, inform animation design by specifying element appearance and narrative focus.  \re{Concrete entity mentions (\eg ``artificial Christmas trees'', ``Americans'', and ``a real one'') are also identified automatically to support the following tasks, such as image filtering and matching.}
\re{The extraction process is conducted through prompt-based queries, and the outputs are further normalized into structured forms. More details can be found in the supplementary materials.}

\textbf{Data Visualization Alternatives.} Based on the uploaded data table and the data-related information extracted from the narrative intents, \re{we expect LLMs to propose feasible designs and provide mapping interfaces to generate charts based on D3.js~\cite{bostock2011d3}}. \re{D3.js enables flexible generation of chart variants, and we provide a set of predefined rendering templates as callable interfaces.} These visualization alternatives are stored in SVG format for subsequent operations and in PNG format \re{to facilitate visual feature extraction by VLMs.}

\textbf{Filtering.} We apply a filtering process to reduce the noise and cognitive load introduced by SAM, while preserving a diverse set of elements for flexible composition.
Semantic-SAM provides coarse-grained labels, allowing the removal of segments unrelated to the narrative theme. \re{Structured shapes are extracted by HED and M-LSD to retain visual distinction.}
Segmented elements are grouped by semantic and contour similarity into unified design components. For instance, in Fig.~\ref{fig: workflow}C, the two trees are merged by recognizing the ``pine tree'' semantics.
The resulting segments, semantic labels, and preprocessing parameters are then passed to the sequential design mapping.

\subsection{Visual Perception}\label{sec: visual perception}
\begin{figure}[h]
  \centering 
  \includegraphics[width=\columnwidth]{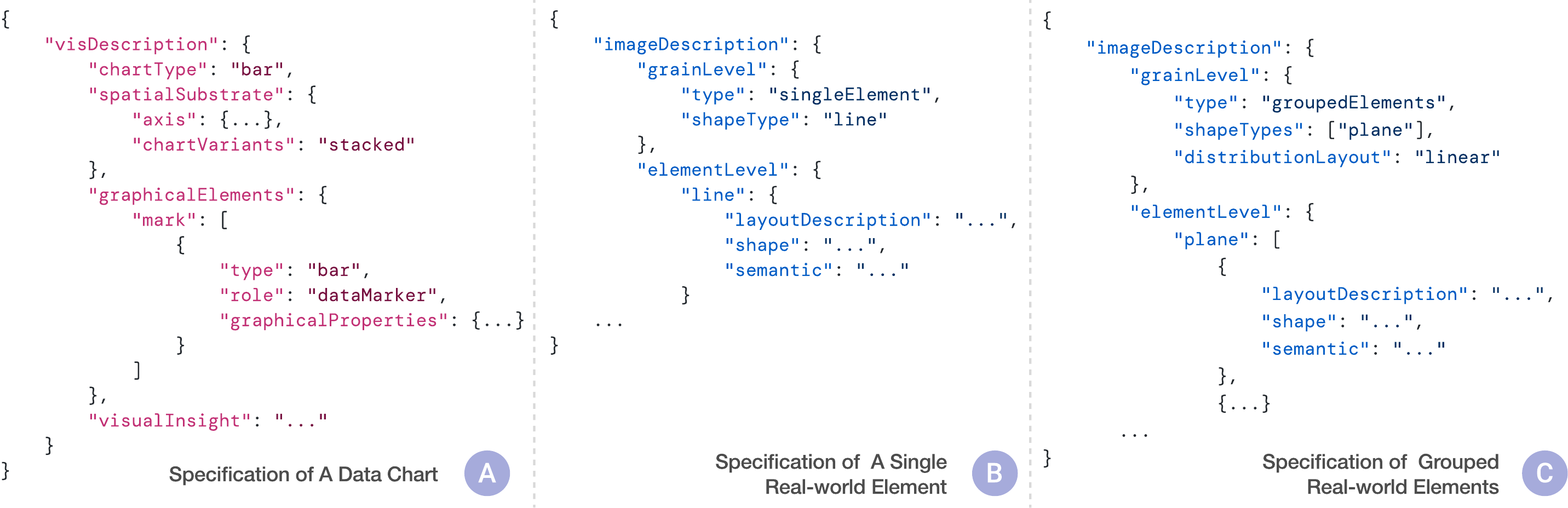}
  \caption{Examples of design component specification. (A) A stacked histogram representing the distribution of tree cover. (B) A single bridge. (C) A combination of two Christmas trees.}
  \label{fig: specification}
\end{figure}

To bridge human perception with AI understanding, we encode key design space dimensions into structured specifications. These specifications help the model reason about each element’s role and provide a consistent reference for subsequent design mapping (Fig.~\ref{fig: specification}).

\textbf{Data Visualization Interpretation and Specification.}
\re{The perception stage inputs both annotated SVGs and corresponding PNGs renderings into the VLMs. The SVGs provide structural and data-specific details, while the PNGs convey visual features such as shape, color, and overall appearance.}
As discussed in Sec.~\ref{sec: vis components}, data visualizations can be interpreted through a three-level visual mapping framework. 
Building on this framework and integrating data semantics, we propose a declarative specification (Fig.~\ref{fig: specification}A) for describing visualization design elements.
\re{The specification begins by defining the chart type, followed by the corresponding visual mappings.
In the spatial substrate section, the data fields are mapped to spatial axes.
To accommodate layout diversity within the same chart type, this level also explicitly specifies the layout variant.
In the graphical elements section, the focus is on geometric shapes used (\eg bar, line, circle) and their associated functional roles (\eg data marker, annotation). Each graphical element is represented as a pair: the element type combined with a natural language description of its encoding channel.
A holistic, insight-driven perspective is adopted when interpreting visualizations. Visual insights are treated as semantic annotations and aggregated visual patterns that support narrative interpretation.}

\textbf{Image Element Understanding and Specification.}
\re{To perceive image design components, the input includes the original image, extracted element masks, and preprocessed features such as detected lines and shapes. As shown in Fig.~\ref{fig: specification}B and C, the specification describes design elements along two dimensions: granularity, distinguishing between individual and grouped elements; and element-level features, capturing geometric and semantic properties. For grouped elements, spatial arrangement and collective geometry are explicitly encoded (Fig.~\ref{fig: specification}C). At the element level, each component is classified by geometric type, layout pattern, and semantic role.}


\subsection{Reasoning and Mapping} \label{sec: reasoning and mapping}
The mapping process is image-driven. Given an image inspiration, VLMs explore possible data visualizations through mapping reasoning (Fig.~\ref{fig: workflow}G–I). This involves design-based reasoning, visualization adjustments, image processing, and evaluation of design alternatives.

\subsubsection{Design Mapping}
Since the design space is organized from a bottom-up perspective, providing guiding principles for VLM understanding helps establish meaningful relationships between fundamental elements. Therefore, we introduce four key considerations \re{through structured prompts (\textit{spatial organization}, \textit{shape similarity}, \textit{layout consistency}, and \textit{semantic binding})} to guide the construction of mapping relationships. \re{By prompting with these considerations, VLMs are encouraged to reason in a Chain-of-Thought (CoT) manner~\cite{jason2022cot}, which is critical for decomposing complex alignment tasks into interpretable, step-by-step decisions.}

First, \textit{\textbf{spatial organization} determines how elements in the scene correspond to visual marks.} A single real-world object may map to an individual data point or a group of related points, while the overall scene may serve as the canvas or coordinate system. For grouped elements, a clear data-binding relationship is essential for meaningful mapping. 
\textit{\textbf{Shape similarity} plays a crucial role in making visualizations intuitive.} Real-world objects should resemble the shapes used in data marks, such as lines in the scene matching line charts, and circular objects aligning with pie charts. Beyond basic shapes, finer details of shape features can also reflect data attributes.  
\textit{\textbf{Layout consistency} ensures that the arrangement of elements in the scene mirrors the visualization structure.} The relative positions of individual objects should align with key points in the visualization, such as axes or reference lines. Meanwhile, the overall distribution of grouped elements (\eg scattered, clustered,~\etc) should match patterns in the data to maintain a coherent spatial relationship.  
Finally, \textit{\textbf{semantic binding} ties meaning to visualization.} Real-world objects should carry direct or metaphorical significance, linking their inherent qualities to data values or categories. Narrative elements in the scene, such as symbolic objects or contextual details, can further enhance this connection, making the visualization accurate and engaging. 
Additionally, an effective design plan should balance semantic relevance and visual alignment. While multiple forms of visual alignment are desirable, they are not strictly required to coexist within a single plan. The model is also expected to suggest potential improvements to better fulfill the intended design goals.

\subsubsection{Visualization Adjustment}
To ensure design coherence while minimizing changes to the original image, we constrain the model to adjust only the visualization. These adjustments preserve the underlying data and operate at two levels.

\textbf{Data-level Adjustment.} The model may perform data binding operations to better align image elements with corresponding data markers by \textit{filtering} data irrelevant to the narrative. For instance, as illustrated in Fig.~\ref{fig: workflow}G, the model removes data entries such as ``No tree'' and ``Don't know'' to better match the remaining values with the two Christmas tree objects. In cases where prominent data insights are present, the model may further apply \textit{classification} or \textit{sorting} strategies to enhance the clarity of visual correspondence. The line chart in Fig.~\ref{fig: system}B3 sorts the data to fit the contour of the sky. \re{The LLM receives both the dataset and a predefined code template, which allows it to perform data transformations and modify the visualization generation.}

\textbf{View-level Adjustment.} This part is intended for visual alignment with the image, requiring operations such as \textit{scale}, \textit{translation}, or \textit{rotation}. Individual visualization elements (\eg a data marker) or the entire visualization are processed as graphical objects. These operations rely on specific image processing parameters, which will be detailed in the next section.

\subsubsection{Tool Execution}
Inspired by Wang~\etal~\cite{wang2024genartist}, the model generates not only design and adjustment strategies but also corresponding tool interfaces and implementation parameters. \re{These are specified through structured prompts and passed internally between system components to support automated operations.} The relevant tools and parameters are listed in \re{the Appendix}. They are primarily designed for manipulating SVG elements within the visualization, including accessing specific elements, managing hierarchical relationships, adjusting element size, position, and rotation angle, and aligning these elements with counterparts in real-world scenes based on various alignment strategies. In addition, all image processing parameters(\eg bounding boxes, anchoring points, rotation angles, ~\etc) are accessible to facilitate tool execution.

\subsubsection{Design Evaluation}
\begin{figure*}[t]
  \centering 
  \includegraphics[width=\linewidth]{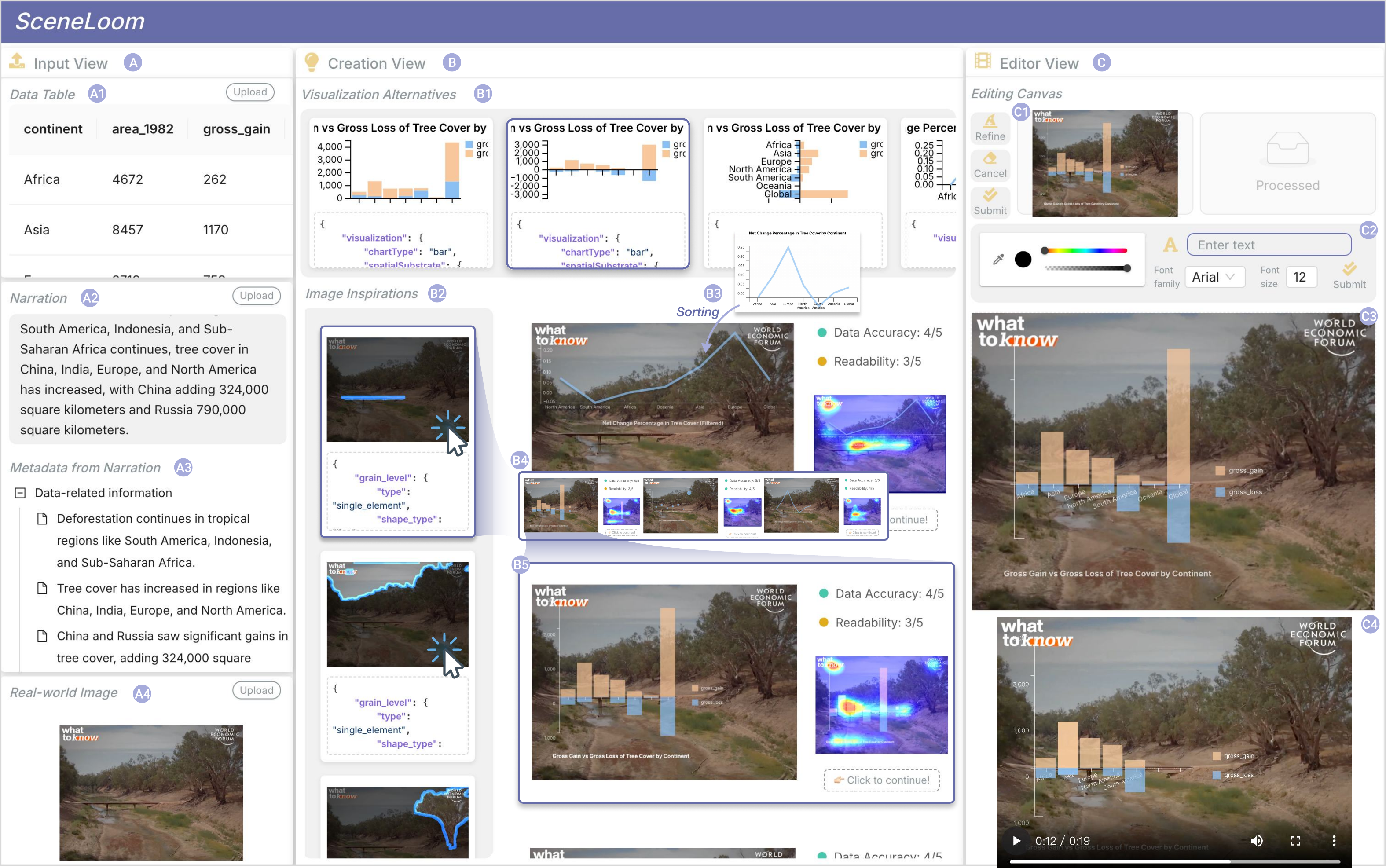}
  \caption{SceneLoom interface within the example of global tree cover change implemented in our user study. After uploading the raw materials (A), users receive a series of inspirations from data visualizations and images. The system supports image-driven browsing and exploration of multiple design alternatives (B). After selecting a design alternative, the system supports fine-tuning and animation generation (C).}
  \label{fig: system}
  \vspace{-0.4cm}
\end{figure*}

\re{The system evaluates design alternatives from data and visual perspectives, presenting data accuracy, visual readability, and attention analysis to assist user refinement.}
For data optimization, it not only verifies accuracy against data tables and bindings, but also detects and resolves conflicts such as encoding inconsistencies and misalignments between data and visual elements.
For example, Fig.\ref{fig: system}J shows a height-encoding conflict, while Fig.\ref{fig: teaser}B highlights mismatches like a hat or hand misaligned with employment data.
These issues are addressed through inpainting~\cite{yu2023inpaint}, reusing and repositioning content when possible, or using semantic-guided generation when necessary.
Visual optimization consists of two components: visual readability and attention analysis. Readability is evaluated qualitatively, with the LLM providing feedback on factors such as the presence of occlusion, color distinction, and layout clarity~\re{, which are presented to users to support design decisions}. Attention analysis uses saliency maps~\cite{sitzmann2018saliency, hosseini2025sum} to simulate eye-tracking and identify visually prominent regions.

\subsection{Interface} \label{sec: interface}
In Fig.~\ref{fig: system}, the interactive interface of SceneLoom comprises three views: \textit{Input View} for data entry, \textit{Creation View} for presenting and selecting design solutions, and \textit{Editor View} for refining and rendering final output.

\textbf{Input View.} The left panel (Fig.~\ref{fig: system}A) enables users to upload multimodal input materials, including data tables (Fig.~\ref{fig: system}A1), narrative intents (Fig.~\ref{fig: system}A2), and real-world images (Fig.~\ref{fig: system}A4). Once the data tables and narrative intents are uploaded, the system extracts relevant information and displays it using a structured, tree-view interface (Fig.~\ref{fig: system}A3).

\textbf{Creation View.} Inspired by VIZITCARDS~\cite{he2017vizitcards}, all design elements are presented as design cards. A diverse collection of data visualization cards constitutes the \textit{Visualization Alternatives} (Fig.~\ref{fig: system}B1), while highlighted visual elements extracted from the image form the \textit{Image Inspirations} (Fig.~\ref{fig: system}B2). These cards are displayed in a scrollable panel layout, with each card encoded using structured specifications aligned with dimensions from our design space. Users are encouraged to explore design possibilities guided by visual cues. All visualization alternatives associated with a selected inspiration are spatially juxtaposed for comparison (Fig.~\ref{fig: system}B4). Each alternative, as illustrated in Fig.~\ref{fig: system}B5, presents a design template and evaluation metrics to support user decision-making, including data expression accuracy, readability, and visual saliency. Upon reviewing the options, users can select a preferred design and proceed by confirming their choice.

\textbf{Editor View.} This view supports the refinement of the selected design alternative and the creation of animations. Automated tools (Fig.~\ref{fig: system}C1) assist in identifying and resolving conflicts between image content and overlaid data through removal and inpainting operations. This step is optional; users seeking to preserve image authenticity may proceed directly to canvas editing. The canvas editor (Fig.\ref{fig: system}C2) allows for fine-grained control over visual elements, including dragging, color adjustment, scaling, rotation, and text insertion. Once editing is complete, SceneLoom generates an animation based on narrative-intent-driven actions and renders a dynamic visualization (Fig.\ref{fig: system}C3), which can be previewed and downloaded automatically.

\section{Evaluation}
We presented an example gallery and conducted a user study to validate the usability and effectiveness of SceneLoom.
\re{All examples discussed in this section were created by participants during the user study. Detailed evaluation results are provided in the Appendix.}

\begin{figure}[t]
  \centering 
  \includegraphics[width=\columnwidth]{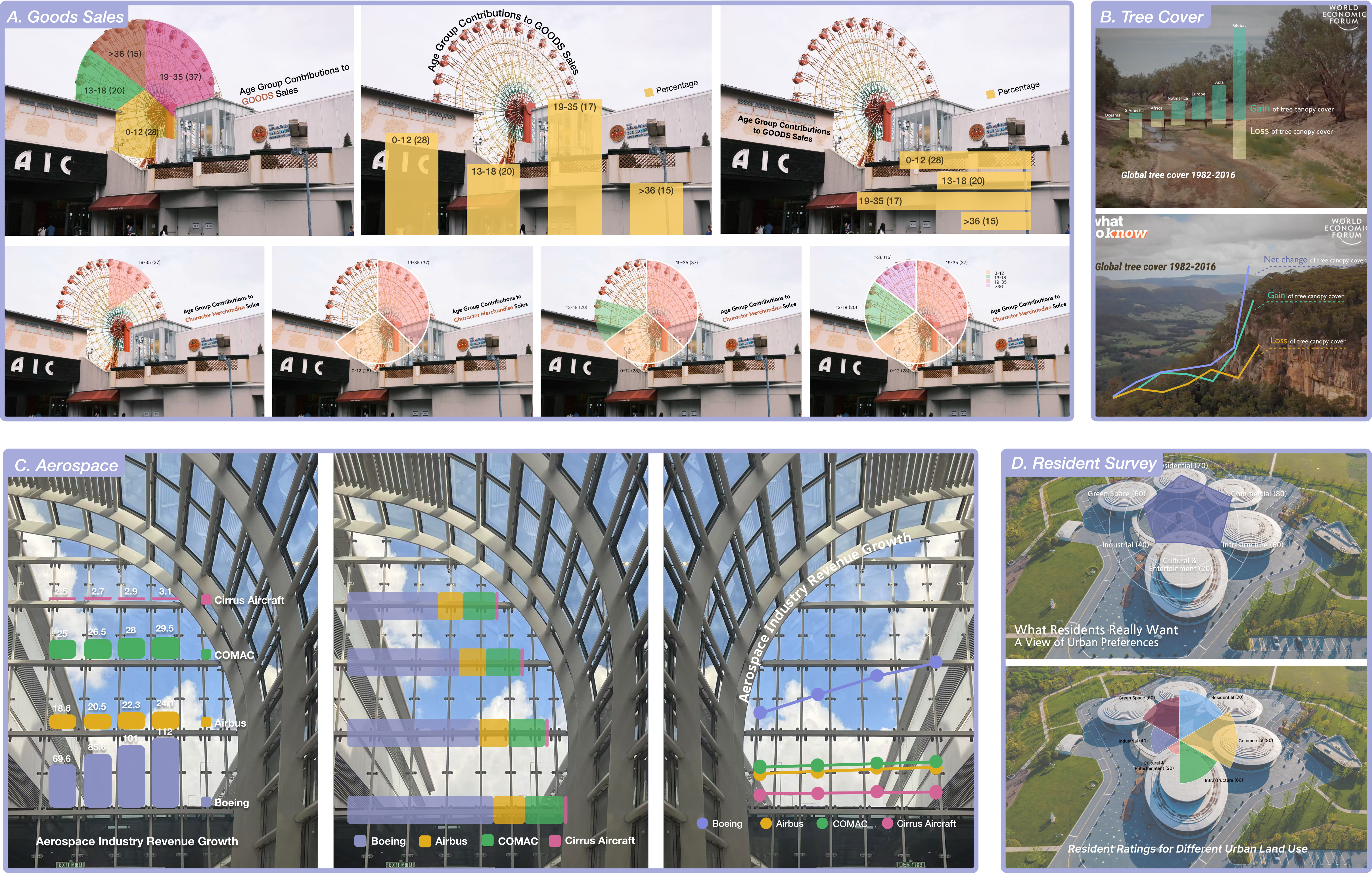}
  \caption{Examples of design outcomes from our user study.
(A) Top: Three designs created by P9 to express the relationship between goods sales and age. Bottom: Animation frames generated by the system.
(B) A composition created by P4 to narrate global tree cover change using imagery from multiple sources.
(C) A design by P8 illustrating the quarterly revenues of four airline companies.
(D) A visualization created by P7 reflecting feedback on residents’ preferences.
\re{These design examples are included in the supplementary materials.}}
  \label{fig: gallery}
\end{figure}

\subsection{Example Gallery}
To demonstrate the expressiveness and appeal of the outcomes produced by users through SceneLoom, we collected design artifacts from participants in the user study, and a selection of them is shown in Fig.~\ref{fig: gallery}. Additional design outcomes, along with their animated versions, are provided in the supplementary materials. The example gallery includes \re{different data themes}, including economic, social, cultural,~\etc We also present different design solutions based on the same materials (Fig.~\ref{fig: gallery}A, C), as well as design solutions generated from the same \re{data table and narrative intents} but with different image inputs (Fig.~\ref{fig: gallery}B). \re{Additional examples are provided in the supplementary materials.}

\subsection{User Study}
\subsubsection{Experimental Set-up}
\textbf{Participants.} We recruited 10 participants (P1-P10) interested in SceneLoom from various fields, including data analysts, graphic designers, journalists, and researchers in HCI or VIS. They were all between 20 and 35, including 6 females and 4 males. We first collected their basic information, through self-report (1 = No experience, 5 = Expert), to understand their expertise in data visualization ($M$=3.80, $SD$=1.03), visual design ($M$=3.40, $SD$=0.84), and video editing ($M$=3.40, $SD$=1.17). Each participant received a \$30 gift card after the study.

\textbf{Procedure.} By presenting several representative examples from the corpus, we introduced the purpose and procedure of the study. Each participant participated in an individual, in-person session, during which they were encouraged to adopt a think-aloud approach to verbalize their thought process.
Firstly, participants were provided with datasets covering 10 topics and three to four relevant real-world images. They were asked to select two sets as the basis for their creative task, either out of interest in a particular topic or inspired by one of the images.
Participants were then given 5–10 minutes to familiarize themselves with the materials. During this time, we asked them about their initial creative ideas and encouraged them to articulate potential challenges. These ideas were primarily conveyed through verbal descriptions or sketches.
Next, we asked participants to use our system to enable this creative process. We introduced the system's tutorial and helped them use the system with an example. In the whole task, participants would independently use SceneLoom to create and implement the selected set to complete the creation within approximately 30 minutes. We documented their findings and issues during the task and, with their consent, saved the final results. Finally, participants completed an exit questionnaire and participated in a semi-structured interview. Each session lasted 40–50 minutes.

\subsubsection{Results and Analysis}

\begin{figure}[t]
  \centering 
  \includegraphics[width=\columnwidth]{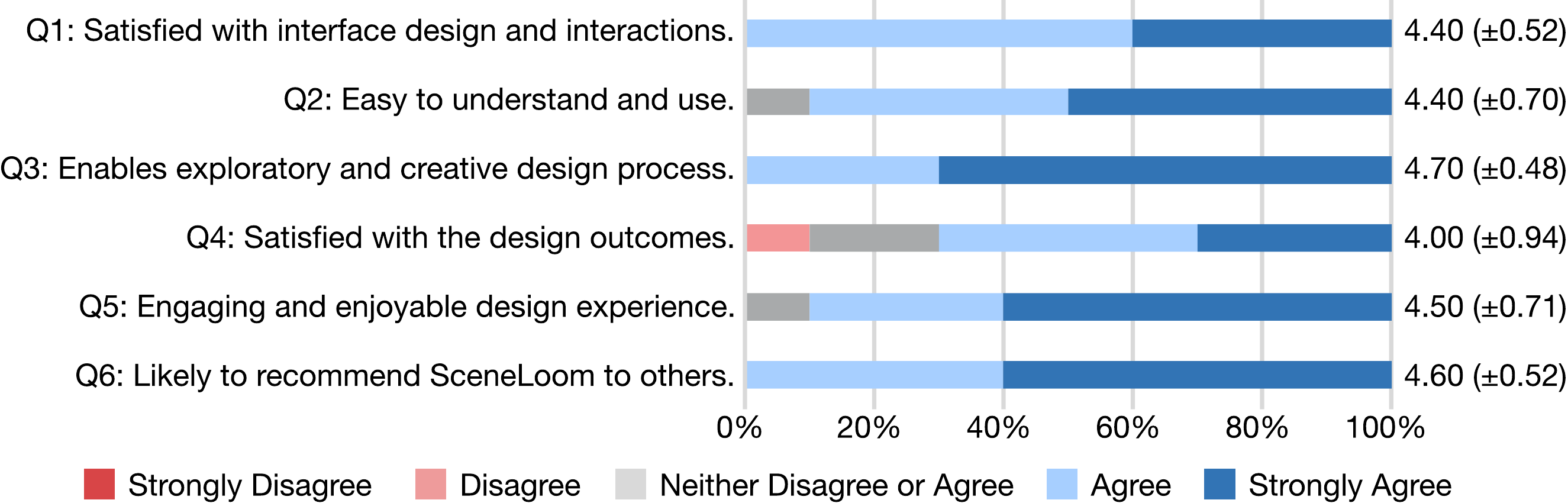}
  \caption{Detailed subjective questions and corresponding user rating results. Q1 and Q2 assess system usability, Q3 and Q4 evaluate effectiveness, and Q5 and Q6 pertain to user recommendations. A 5-point Likert scale was employed to quantify user satisfaction, where a score of 5 represents strong agreement.}
  \label{fig: user study}
  \vspace{-0.45cm}
\end{figure}

\textbf{Quantitative results of the questionnaire.} All users completed the assigned tasks and provided feedback on SceneLoom, as shown in Fig.~\ref{fig: user study}. 
We evaluated the system using a 5-point Likert scale from the perspectives of usability, effectiveness, and recommendability. The system design ($M$=4.40, $SD$=0.52) and user experience ($M$=4.40, $SD$=0.70) received positive feedback. Throughout the experiment, participants expressed a strong desire for exploration and demonstrated rich design ideas ($M$=4.70, $SD$=0.48).
For the presentation of the final results ($M$=4.00, $SD$=0.94), some participants noted that the outcomes required further refinement. However, others reported that the intended design concepts were effectively communicated, and the need for post-adjustments did not significantly affect their overall perception.
The overall design experience received positive feedback ($M$=4.50, $SD$=0.71), and participants were willing to recommend SceneLoom to others or use it in the future ($M$=4.60, $SD$=0.52).

\textbf{The overall workflow effectively facilitated the generation and refinement of creative designs.} Participants reported that SceneLoom provided substantial support at each stage of our workflow.
P9 praised the effectiveness of data preparation. She remarked regarding the extraction of narrative intent, \textit{``This structured representation of narrative intent allowed me to verify key information, avoiding omissions. It also enabled me to stay narrative-driven throughout the design process.''} 
Participants emphasized the importance of integrating both data visualizations and real-world imagery during feature extraction. Some (4/10) reported that they often relied solely on one modality, either due to data complexity or visual bias, which sometimes resulted in missed design opportunities. P7 noted that the abstract explanations increased cognitive load, suggesting that natural language guidance might offer clearer support. Conversely, P2 appreciated the structured form, noting that \textit{``it allowed me to align design suggestions with underlying principles, thereby enhancing the credibility of the generated outputs''}. This feedback highlights the need for more adaptive explanation strategies.
Participants responded positively to the diversity and creativity of the design drafts produced by SceneLoom. These outputs validated initial ideas and inspired new directions, making the design process more exploratory. P4 praised the system’s adaptability, sharing, \textit{``I hadn't thought about reordering the data to match the contour features in the image… I’ll definitely use this again to resolve mismatches.''}
Overall, participants expressed high satisfaction with the outcomes. P8 remarked that \textit{``SceneLoom's recommendations exceeded my original ideas.''} The integration of animation further contributed to a cohesive and polished presentation. Although P9 initially raised concerns about visual occlusion, she later found that the sequential animation effectively balanced content visibility with narrative flow (see Fig.~\ref{fig: gallery}A).

\textbf{Design externalizations support creative ideation and assist in evaluating, selecting, and refining design solutions.} Among the participants, some (2/10) had limited experience with data visualization, and some (3/10) were less proficient in extracting visual inspiration from images. 
After using SceneLoom, most participants (8/10) remarked that having abstract data represented through diverse visualizations allowed them to intuitively grasp distribution patterns and spatial configurations, \re{as the visual forms made even simple datasets more interpretable and revealing.}
As P6 noted, \textit{``Without the data visualization alternatives, I felt lost in the design process and could only form a vague idea.''} \re{Multi-faceted data often makes it challenging to establish meaningful connections to visual design. For instance, one of the examples featured quarterly production data from four aircraft manufacturers. P2 said, \textit{``Even though the aircraft production data only had a few columns, I found it hard to see the temporal trends or differences between brands just by looking at numbers. SceneLoom enabled me to better comprehend how visual forms can represent such data more clearly.''}}
Moreover, presenting the design draft gave participants useful references for making informed choices, evaluating alternatives, and iterating their ideas. For instance, in the case of economic goods (Fig.~\ref{fig: gallery}A), P9 initially intended to use railings as a metaphor for data mapping. However, after viewing the generated design inspired by railings, she reflected, \textit{``The combination didn’t work as well as I had expected -- the visualization appeared abrupt in the scene.''} This led her to reconsider and ultimately discard that design direction.
Similarly, P5, who chose to represent a pie chart using the Ferris wheel structure, realized that its placement interfered with foreground buildings. Drawing on SceneLoom’s readability suggestions, she refined the layout by reusing previously extracted building elements and layering them over the pie chart, thereby preserving the original visual depth of the scene.

\textbf{Diverse design solutions reduced cognitive fixation and provided practical options for different storytelling needs.} Most of the participants (9/10) reported that they often fell into fixed design patterns when encountering data or images, limiting their creative thinking. This rigidity was partly due to an overemphasis on dominant visual elements. For example, P1 remarked, \textit{``I was too focused on the Ferris wheel and overlooked the surrounding details. I liked this design—it felt like a subtle but clever idea and brought an element of surprise.''} \re{P8 echoed this view, noting that SceneLoom’s extraction of diverse visual features helped them move beyond dominant elements and discover overlooked but meaningful details.}
Design fixation was also evident in participants' visualization preferences. As P9 explained, \textit{``If not for the diverse visualization choices, I probably would have defaulted to traditional bar charts without considering alternative layouts.''} She continued to realize that such variation can significantly impact how the visualization integrates with the real-world scene.
P4 noted that SceneLoom effectively mapped data to relevant objects using insightful, context-aware strategies in the tree cover case (Fig.~\ref{fig: gallery}B), making the process both engaging and valuable. P6 added that design diversity supports different communication goals. For example, some layouts suit formal, fact-based narratives, while others are better for general presentations, education, or storytelling.

\textbf{Analysis of failure cases.} Among the 16 participants, three did not initially receive design suggestions from SceneLoom, and one noticed that an obvious design pattern was not identified. Nonetheless, they remained patient and were willing to try again. We analyzed these cases and summarized the following reasons: (1) Simple narrative intents and insufficient semantic information limit effective filtering and reference for image elements, reducing inspiration. For example, when a user described \textit{``optimistic oil consumption year by year''}, SceneLoom struggled to match this with relevant elements as no ``oil'' exists in the image. (2) High image complexity~\re{, characterized by dense visual elements, layered spatial structures, and intricate textures (\eg urban streetscape, busy indoor scenes), complicates segmentation and overlay alignment.} Conversely, simple images also made it hard to identify suitable objects. In such cases, we suggested users either replace the image or apply a basic overlay approach.
Apart from missing content, two design errors resulted from misalignment. While the design plans were correct and valid, irregular bounding boxes and occlusions caused positioning deviations, requiring manual adjustments. These common computer vision challenges were amplified in our scenario, which demanded both semantic and visual accuracy.
Future work could further enhance the natural language understanding and object detection abilities of the system by tracking the most advanced models.

\section{Discussion}\label{sec: discuss}
In this session, we revisit our coordination method and system to discuss current limitations and key opportunities for advancing its adaptability,  expressiveness, and user engagement.

\textbf{Towards more expressive coordination between data visualization and real-world contexts.} Our work presents a foundational approach for coordination while maintaining visual consistency and semantic coherence. 
\re{While our corpus analysis focused on video-based storytelling, discussions with experts and users suggested that the proposed design space may generalize to other formats, such as interactive articles and scrollytelling. Future work will explore its adaptability across media to assess broader applicability and medium-specific considerations.}
\re{Layout strategies like separation with curtain-opening effect, juxtaposition, and background substitution enable diverse narrative expressions. Each entails unique design and technical demands, from scene segmentation to rendering workflows.}
In narrative-driven contexts, additional factors often shape coordination. Aesthetic principles~\cite{kong2023aethesics} such as contrast, hierarchy, and balance can inform the evaluation and refinement of design alternatives. Emotional cues~\cite{lan2024affective} also play a crucial role in enhancing audience engagement and may guide the mapping process. 
We also observed that different coordination strategies influence the type and timing of animations, opening opportunities for motion design guided by narrative intent. Looking ahead, we envision extending this coordination framework to AR environments, which offer more immersive and spatially anchored storytelling~\cite{willett2017embedded}.

\textbf{Toward a more flexible and scalable workflow.} Our current workflow requires users to provide diverse and multimodal input data. While this enables a clearer understanding of design requirements, it also introduces considerable challenges in data collection and preprocessing. To address this, we envision extending the workflow to support more generalized and intuitive input forms. For instance, users could provide a real-world video as input, from which the system could automatically extract relevant keyframes for subsequent coordination. Alternatively, generative model-based methods could be integrated to support user-driven creation directly from semantic-level inputs.
On the output side, we also see potential for diversification. Our current implementation presents the output as a dynamic video clip. Depending on the context and narrative goals, this can be extended to support a wider range of storytelling scenarios like infographics or data news.

\textbf{Supporting a more customized and mixed-initiative creation process.} Our user study revealed significant diversity in user preferences related to visual design. Some users prioritized the clear presentation of data, favoring representations grounded in statistical accuracy, while others preferred more imaginative and expressive visual forms. For some, semantic coherence was critical, whereas others were more drawn to visual aesthetics and emotional appeal.
Incorporating user preferences as conditioning factors within the model's reasoning and decision-making processes is crucial to accommodate this diversity. This enables a more personalized design experience while fostering effective human-AI collaboration~\cite{Instructions,shen2024dataplaywright, NotePlayer}. Such a \re{human-in-the-loop} approach also helps navigate the trade-off between data fidelity and creative flexibility.
Furthermore, introducing direct or hybrid interaction methods, such as selecting, linking, and direct manipulation of visual elements, can make the creation process more intuitive~\cite{wonderflow}.

\textbf{LLM performance for creative support.} The flexibility of LLMs makes them well-suited for creative support tasks~\cite{NarrativePlayer,datadirector,dataplayer}. However, in our current workflow, we have observed several challenges that affect their reliability, efficiency, and creative diversity. Due to the complexity of task-specific reasoning and multimodal inputs, the response time of LLMs can vary.
\re{Each case takes on average 77.6 seconds, with 13,803 input and 1,832 output tokens required to generate a single result. While this cost is generally acceptable, handling multiple candidates introduces moderate additional overhead, highlighting the importance of efficient pruning and scheduling.}
Such latency can affect the system’s ability to provide timely feedback, which is critical for maintaining user engagement and creative flow. Although presenting intermediate reasoning steps in the interface offers valuable insights, users still encounter inconsistent waiting periods. To improve responsiveness, we can explore strategies such as optimizing reasoning pipelines and adopting asynchronous response mechanisms~\cite{ginart2024asynchronoustoolusagerealtime}.
Another challenge lies in the model’s limited familiarity with tool usage. Despite efforts to standardize tool descriptions and prompt design, the model may still hallucinate unsupported functions in complex scenarios. This suggests the need to further expand the tool library~\cite{cai2024largelanguagemodelstool} and explore fine-tuning approaches~\cite{gao2025fine-tuned}.
Additionally, while we incorporate design knowledge to support more grounded and comprehensive suggestions, the model still exhibits preferences. Understanding and mitigating such biases remains an important direction for future work.

\textbf{Limitations and future work.} Our evaluation strategy primarily relies on user studies to assess users' experiences with the system and their satisfaction with the resulting designs. Although SceneLoom received encouraging feedback in this limited-scale study, we recognize the need to broaden the participant base in future work. 
\re{Involving a more diverse user base and collecting richer feedback enables iterative refinement of the system and its design space.}
Moreover, current forms of data visualization are mainly based on basic chart types and are refined through simple modifications, which limits their expressive power. In future work, we aim to explore more creative and visually compelling forms of data visualization. One promising direction involves leveraging generative models to synthesize entire visualizations and accompanying elements such as icons and illustrations.
Furthermore, as the creative process becomes increasingly collaborative~\cite{lee2021shared}, we envision supporting co-creation and content sharing within the system. For instance, a community-oriented platform, like Pinterest, could be developed to facilitate the exchange of ideas, design inspiration, and peer feedback. Such a platform would further foster creativity through social interaction and collective refinement of visual concepts.







\section{Conclusion}
This paper presents SceneLoom, a VLM-powered system that enables the coordinated integration of data visualizations and real-world imagery for expressive data storytelling. Grounded in a formative study, we identified key design considerations. SceneLoom uses VLMs to extract and reason over these features, generating diverse, contextually aligned design alternatives that support narrative intent. Users can explore, refine, and animate these designs to externalize ideas and enhance visual communication. Our user study and example gallery validate SceneLoom’s ability to inspire creativity and expand the expressive potential of narrative visualization.
We hope our approach could help users enhance creative data communication and inspire future work.

\end{spacing}

\bibliographystyle{abbrv-doi-hyperref}

\newpage
\acknowledgments{%
This work is supported by Natural Science Foundation of China (NSFC No.62472099) and Ji Hua Laboratory S\&T Program (X250881UG250). 
We sincerely thank Prof. Xingyu Lan for her valuable feedback.
We also thank my friend Liwenhan Xie at HKUST for the insightful discussions related to this work.
%
}

\bibliography{main}

\begin{thebibliography}{10}

\bibitem{achiam2023gpt}
J.~Achiam, S.~Adler, S.~Agarwal, L.~Ahmad, I.~Akkaya, F.~L. Aleman, D.~Almeida, J.~Altenschmidt, S.~Altman, S.~Anadkat, et~al.
\newblock Gpt-4 technical report.
\newblock {\em arXiv preprint arXiv:2303.08774}, 2023.

\bibitem{avrahami2022inpainting}
O.~Avrahami, D.~Lischinski, and O.~Fried.
\newblock Blended diffusion for text-driven editing of natural images.
\newblock In {\em 2022 IEEE/CVF Conference on Computer Vision and Pattern Recognition (CVPR)}, vol. 2022, pp. 18187--18197, 2022.

\bibitem{bostock2011d3}
M.~Bostock, V.~Ogievetsky, and J.~Heer.
\newblock D³ data-driven documents.
\newblock {\em IEEE Trans. Vis. Comput. Graph.}, 17(12):2301--2309, 2011.

\bibitem{cai2024largelanguagemodelstool}
T.~Cai, X.~Wang, T.~Ma, X.~Chen, and D.~Zhou.
\newblock Large language models as tool makers.
\newblock In {\em Proc. ICLR}, 2024.

\bibitem{card1999readings}
S.~K. Card, J.~Mackinlay, and B.~Shneiderman.
\newblock {\em Readings in information visualization: using vision to think}.
\newblock Morgan Kaufmann, 1999.

\bibitem{chen2024beyond}
Q.~Chen, W.~Shuai, J.~Zhang, Z.~Sun, and N.~Cao.
\newblock Beyond numbers: Creating analogies to enhance data comprehension and communication with generative ai.
\newblock In {\em Proceedings of the 2024 CHI Conference on Human Factors in Computing Systems}, CHI '24. ACM, 2024.

\bibitem{cheng2023prior}
P.~Cheng, L.~Lin, J.~Lyu, Y.~Huang, W.~Luo, and X.~Tang.
\newblock Prior: Prototype representation joint learning from medical images and reports.
\newblock In {\em 2023 IEEE/CVF International Conference on Computer Vision (ICCV)}, vol. 2023, pp. 21304--21314, 2023.

\bibitem{chilton2021visifit}
L.~B. Chilton, E.~J. Ozmen, S.~H. Ross, and V.~Liu.
\newblock Visifit: Structuring iterative improvement for novice designers.
\newblock In {\em Proceedings of the 2021 CHI Conference on Human Factors in Computing Systems}. ACM, 2021.

\bibitem{chilton2019visiblends}
L.~B. Chilton, S.~Petridis, and M.~Agrawala.
\newblock Visiblends: A flexible workflow for visual blends.
\newblock In {\em Proceedings of the 2019 CHI Conference on Human Factors in Computing Systems}, CHI '19, p. 1–14. ACM, 2019.

\bibitem{choi2024creative}
D.~Choi, S.~Hong, J.~Park, J.~J.~Y. Chung, and J.~Kim.
\newblock Creativeconnect: Supporting reference recombination for graphic design ideation with generative ai.
\newblock In {\em Proceedings of the 2024 CHI Conference on Human Factors in Computing Systems}, CHI '24. ACM, 2024.

\bibitem{coelho202infomage}
D.~Coelho and K.~Mueller.
\newblock {Infomages: Embedding Data into Thematic Images}.
\newblock {\em Computer Graphics Forum}, 2020.

\bibitem{gao2025fine-tuned}
L.~Gao, J.~Lu, Z.~Shao, Z.~Lin, S.~Yue, C.~Leong, Y.~Sun, R.~J. Zauner, Z.~Wei, and S.~Chen.
\newblock Fine-tuned large language model for visualization system: A study on self-regulated learning in education.
\newblock {\em IEEE Trans. Vis. Comput. Graph.}, 31(1):514--524, 2025.

\bibitem{ginart2024asynchronoustoolusagerealtime}
A.~A. Ginart, N.~Kodali, J.~Lee, C.~Xiong, S.~Savarese, and J.~Emmons.
\newblock Asynchronous tool usage for real-time agents.
\newblock {\em arXiv preprint arXiv:2410.21620}, 2024.

\bibitem{gu2022mlsd}
G.~Gu, B.~Ko, S.~Go, S.-H. Lee, J.~Lee, and M.~Shin.
\newblock Towards light-weight and real-time line segment detection.
\newblock {\em Proceedings of the AAAI Conference on Artificial Intelligence}, 36(1):726--734, 2022.

\bibitem{guardian2014uk}
T.~Guardian.
\newblock Edward snowden and the nsa files: facts and figures, 2014.
\newblock \url{https://www.youtube.com/watch?v=OFCNqkDWMtY&t=110s }. Accessed: 2025-03-23.

\bibitem{guo2024understanding}
G.~Guo, J.~J. Kang, R.~S. Shah, H.~Pfister, and S.~Varma.
\newblock Understanding graphical perception in data visualization through zero-shot prompting of vision-language models.
\newblock {\em arXiv preprint arXiv:2411.00257}, 2024.

\bibitem{Guo2024regiongpt}
Q.~Guo, S.~De~Mello, H.~Yin, W.~Byeon, K.~C. Cheung, Y.~Yu, P.~Luo, and S.~Liu.
\newblock Regiongpt: Towards region understanding vision language model.
\newblock In {\em Proceedings of the IEEE/CVF Conference on Computer Vision and Pattern Recognition (CVPR)}, pp. 13796--13806, 2024.

\bibitem{he2017vizitcards}
S.~He and E.~Adar.
\newblock Vizitcards: A card-based toolkit for infovis design education.
\newblock {\em IEEE Trans. Vis. Comput. Graph.}, 23(1):561--570, 2017.

\bibitem{herman2025touch}
B.~Herman, C.~D. Jackson, and D.~F. Keefe.
\newblock Touching the ground: Evaluating the effectiveness of data physicalizations for spatial data analysis tasks.
\newblock {\em IEEE Trans. Vis. Comput. Graph.}, 31(1):875--885, 2025.

\bibitem{hosseini2025sum}
A.~Hosseini, A.~Kazerouni, S.~Akhavan, M.~Brudno, and B.~Taati.
\newblock Sum: Saliency unification through mamba for visual attention modeling.
\newblock In {\em Proc. WACV}, pp. 1597--1607, 2025.

\bibitem{huang2015personal}
D.~Huang, M.~Tory, B.~Adriel~Aseniero, L.~Bartram, S.~Bateman, S.~Carpendale, A.~Tang, and R.~Woodbury.
\newblock Personal visualization and personal visual analytics.
\newblock {\em IEEE Trans. Vis. Comput. Graph.}, 21(3):420--433, 2015.

\bibitem{islam2024lvlms}
M.~S. Islam, R.~Rahman, A.~Masry, M.~T.~R. Laskar, M.~T. Nayeem, and E.~Hoque.
\newblock Are large vision language models up to the challenge of chart comprehension and reasoning.
\newblock In {\em Findings of EMNLP 2024}, pp. 3334--3368. ACL, 2024.

\bibitem{wsj2024haidilao}
T.~W.~S. Journal.
\newblock This chinese restaurant chain built its \$9b empire off customer service, 2024.
\newblock \url{https://www.youtube.com/watch?v=0jci98uOrWQ&t=170s}. Accessed: 2025-03-23.

\bibitem{kang2021metamap}
Y.~Kang, Z.~Sun, S.~Wang, Z.~Huang, Z.~Wu, and X.~Ma.
\newblock Metamap: Supporting visual metaphor ideation through multi-dimensional example-based exploration.
\newblock In {\em Proceedings of the 2021 CHI Conference on Human Factors in Computing Systems}, CHI '21. ACM, 2021.

\bibitem{kirillov2023sam}
A.~Kirillov, E.~Mintun, N.~Ravi, H.~Mao, C.~Rolland, L.~Gustafson, T.~Xiao, S.~Whitehead, A.~C. Berg, W.-Y. Lo, P.~Dollár, and R.~Girshick.
\newblock Segment anything.
\newblock In {\em 2023 IEEE/CVF International Conference on Computer Vision (ICCV)}, vol. 2023, pp. 3992--4003, 2023.

\bibitem{kong2023aethesics}
W.~Kong, Z.~Jiang, S.~Sun, Z.~Guo, W.~Cui, T.~Liu, J.~Lou, and D.~Zhang.
\newblock Aesthetics++: Refining graphic designs by exploring design principles and human preference.
\newblock {\em IEEE Trans. Vis. Comput. Graph.}, 29(6):3093–3104, 2023.

\bibitem{kouts2023lsdvis}
A.~Kouts, L.~Besan{\c{c}}on, M.~Sedlmair, and B.~Lee.
\newblock Lsdvis: Hallucinatory data visualisations in real world environments.
\newblock {\em arXiv preprint arXiv:2312.11144}, 2023.

\bibitem{kuruvilla2016review}
J.~Kuruvilla, D.~Sukumaran, A.~Sankar, and S.~P. Joy.
\newblock A review on image processing and image segmentation.
\newblock In {\em 2016 International Conference on Data Mining and Advanced Computing (SAPIENCE)}, pp. 198--203, 2016.

\bibitem{simon2016ideation}
S.~Laing and M.~Masoodian.
\newblock A study of the influence of visual imagery on graphic design ideation.
\newblock {\em Design Studies}, 45:187--209, 2016.

\bibitem{lan2024affective}
X.~Lan, Y.~Wu, and N.~Cao.
\newblock Affective visualization design: Leveraging the emotional impact of data.
\newblock {\em IEEE Trans. Vis. Comput. Graph.}, 30(1):1--11, 2024.

\bibitem{lee2021shared}
B.~Lee, X.~Hu, M.~Cordeil, A.~Prouzeau, B.~Jenny, and T.~Dwyer.
\newblock Shared surfaces and spaces: Collaborative data visualisation in a co-located immersive environment.
\newblock {\em IEEE Trans. Vis. Comput. Graph.}, 27(2):1171--1181, 2021.

\bibitem{li2024semanticsam}
F.~Li, H.~Zhang, P.~Sun, X.~Zou, S.~Liu, C.~Li, J.~Yang, L.~Zhang, and J.~Gao.
\newblock Segment and recognize anything at any granularity.
\newblock In {\em 18th European Conference on Computer Vision, ECCV'24}, p. 467–484. Springer-Verlag, 2024.

\bibitem{li2025composing}
H.~Li, L.~Ying, L.~Shen, Y.~Wang, Y.~Wu, and H.~Qu.
\newblock {Composing Data Stories with Meta Relations}.
\newblock {\em arXiv preprint arXiv:2501.03603}, 2025.

\bibitem{li2024situationadapt}
Z.~Li, C.~Gebhardt, Y.~Inglin, N.~Steck, P.~Streli, and C.~Holz.
\newblock Situationadapt: Contextual ui optimization in mixed reality with situation awareness via llm reasoning.
\newblock In {\em Proceedings of the 37th Annual ACM Symposium on User Interface Software and Technology}, UIST '24. ACM, 2024.

\bibitem{liu2023visual}
H.~Liu, C.~Li, Q.~Wu, and Y.~J. Lee.
\newblock Visual instruction tuning.
\newblock In {\em Thirty-seventh Conference on Neural Information Processing Systems}, 2023.

\bibitem{liu2025smartboard}
Z.~Liu, X.~Xie, M.~He, W.~Zhao, Y.~Wu, L.~Cheng, H.~Zhang, and Y.~Wu.
\newblock Smartboard: Visual exploration of team tactics with llm agent.
\newblock {\em IEEE Trans. Vis. Comput. Graph.}, 31(1):23--33, 2025.

\bibitem{lukac2013painting}
M.~Luk\'{a}\v{c}, J.~Fi\v{s}er, J.-C. Bazin, O.~Jamri\v{s}ka, A.~Sorkine-Hornung, and D.~S\'{y}kora.
\newblock Painting by feature: texture boundaries for example-based image creation.
\newblock {\em ACM Trans. Graph.}, 32(4),  article no. 116, 2013.

\bibitem{lundgard2022accessible}
A.~Lundgard and A.~Satyanarayan.
\newblock Accessible visualization via natural language descriptions: A four-level model of semantic content.
\newblock {\em IEEE Trans. Vis. Comput. Graph.}, 28(1):1073--1083, 2022.

\bibitem{mac2024basketball}
M.~MacKelvie.
\newblock The clutch goat...(it's not who you think).
\newblock \url{https://www.youtube.com/watch?v=qjjW1l9KjXQ&t=763s }. Accessed: 2025-03-23.

\bibitem{meng2025mmiu}
F.~Meng, J.~Wang, C.~Li, Q.~Lu, H.~Tian, T.~Yang, J.~Liao, X.~Zhu, J.~Dai, Y.~Qiao, P.~Luo, K.~Zhang, and W.~Shao.
\newblock {MMIU}: Multimodal multi-image understanding for evaluating large vision-language models.
\newblock In {\em Proc. ICLR}, 2025.

\bibitem{morais2022showing}
L.~Morais, Y.~Jansen, N.~Andrade, and P.~Dragicevic.
\newblock Showing data about people: A design space of anthropographics.
\newblock {\em IEEE Trans. Vis. Comput. Graph.}, 28(3):1661--1679, 2022.

\bibitem{NotePlayer}
Y.~Ouyang, L.~Shen, Y.~Wang, and Q.~Li.
\newblock {NotePlayer: Engaging Computational Notebooks for Dynamic Presentation of Analytical Processes}.
\newblock In {\em Proceedings of the 37th Annual ACM Symposium on User Interface Software and Technology}, pp. 1--20. ACM, 2024.

\bibitem{rombach2022stable}
R.~Rombach, A.~Blattmann, D.~Lorenz, P.~Esser, and B.~Ommer.
\newblock { High-Resolution Image Synthesis with Latent Diffusion Models }.
\newblock pp. 10674--10685. IEEE, 2022.

\bibitem{NarrativePlayer}
Z.~Shao, L.~Shen, H.~Li, Y.~Shan, H.~Qu, Y.~Wang, and S.~Chen.
\newblock {Narrative Player: Reviving Data Narratives with Visuals}.
\newblock {\em IEEE Trans. Vis. Comput. Graph.}, pp. 1--15, 2025.

\bibitem{shen2024dataplaywright}
L.~Shen, H.~Li, Y.~Wang, T.~Luo, Y.~Luo, and H.~Qu.
\newblock Data playwright: Authoring data videos with annotated narration.
\newblock {\em IEEE Trans. Vis. Comput. Graph.}, pp. 1--14, 2024.

\bibitem{datadirector}
L.~Shen, H.~Li, Y.~Wang, and H.~Qu.
\newblock {From Data to Story: Towards Automatic Animated Data Video Creation with LLM-Based Multi-Agent Systems}.
\newblock In {\em IEEE VIS 2024 Workshop on Data Storytelling in an Era of Generative AI, GEN4DS‘24}, pp. 20--27. IEEE, 2024.

\bibitem{shen2025reflecting}
L.~Shen, H.~Li, Y.~Wang, and H.~Qu.
\newblock {Reflecting on Design Paradigms of Animated Data Video Tools}.
\newblock In {\em Proceedings of the 2025 CHI Conference on Human Factors in Computing Systems}, pp. 1--21. ACM, 2025.

\bibitem{Instructions}
L.~Shen, H.~Li, Y.~Wang, X.~Xie, and H.~Qu.
\newblock {Prompting Generative AI with Interaction-Augmented Instructions}.
\newblock In {\em Extended Abstracts of the CHI Conference on Human Factors in Computing Systems, CHI EA '25}, pp. 1--9. ACM, 2025.

\bibitem{shen2023towards}
L.~Shen, E.~Shen, Y.~Luo, X.~Yang, X.~Hu, X.~Zhang, Z.~Tai, and J.~Wang.
\newblock Towards natural language interfaces for data visualization: A survey.
\newblock {\em IEEE Trans. Vis. Comput. Graph.}, 29(6):3121–3144, 2023.

\bibitem{dataplayer}
L.~Shen, Y.~Zhang, H.~Zhang, and Y.~Wang.
\newblock {Data Player: Automatic Generation of Data Videos with Narration-Animation Interplay}.
\newblock {\em IEEE Trans. Vis. Comput. Graph.}, 30(1):109--119, 2024.

\bibitem{shi2024mondrian}
X.~Shi, M.~Liu, Z.~Zhou, A.~Neshati, R.~Rossi, and J.~Zhao.
\newblock Exploring interactive color palettes for abstraction-driven exploratory image colorization.
\newblock In {\em Proceedings of the 2024 CHI Conference on Human Factors in Computing Systems}, CHI '24. ACM, 2024.

\bibitem{shi2025brickify}
X.~Shi, Y.~Wang, R.~Rossi, and J.~Zhao.
\newblock Brickify: Enabling expressive design intent specification through direct manipulation on design tokens.
\newblock {\em arXiv preprint arXiv:2502.21219}, 2025.

\bibitem{shi2021comunicating}
Y.~Shi, X.~Lan, J.~Li, Z.~Li, and N.~Cao.
\newblock Communicating with motion: A design space for animated visual narratives in data videos.
\newblock In {\em Proceedings of the 2021 CHI Conference on Human Factors in Computing Systems}, CHI '21. ACM, 2021.

\bibitem{sitzmann2018saliency}
V.~Sitzmann, A.~Serrano, A.~Pavel, M.~Agrawala, D.~Gutierrez, B.~Masia, and G.~Wetzstein.
\newblock Saliency in vr: How do people explore virtual environments?
\newblock {\em IEEE Trans. Vis. Comput. Graph.}, 24(4):1633--1642, 2018.

\bibitem{tang2022smartshots}
T.~Tang, J.~Tang, J.~Lai, L.~Ying, Y.~Wu, L.~Yu, and P.~Ren.
\newblock Smartshots: An optimization approach for generating videos with data visualizations embedded.
\newblock {\em ACM Trans. Interact. Intell. Syst.}, 12(1), 2022.

\bibitem{tong2024vistellvr}
W.~Tong, K.~Shigyo, L.-P. Yuan, M.~Fan, T.-C. Pong, H.~Qu, and M.~Xia.
\newblock Vistellar: Embedding data visualization to short-form videos using mobile augmented reality.
\newblock {\em IEEE Trans. Vis. Comput. Graph.}, 31(3):1862--1874, 2025.

\bibitem{vox2023soccer}
Vox.
\newblock World cup penalty kicks, tracked, 2023.
\newblock \url{https://www.youtube.com/watch?v=HAuwPue57Vs&t=151s }. Accessed: 2025-03-23.

\bibitem{wang2025chartinsighter}
F.~Wang, B.~Wang, X.~Shu, Z.~Liu, Z.~Shao, C.~Liu, and S.~Chen.
\newblock Chartinsighter: An approach for mitigating hallucination in time-series chart summary generation with a benchmark dataset.
\newblock {\em arXiv preprint arXiv:2501.09349}, 2025.

\bibitem{wang2023visonllm}
W.~Wang, Z.~Chen, X.~Chen, J.~Wu, X.~Zhu, G.~Zeng, P.~Luo, T.~Lu, J.~Zhou, Y.~Qiao, and J.~Dai.
\newblock Visionllm: large language model is also an open-ended decoder for vision-centric tasks.
\newblock In {\em Proceedings of the 37th International Conference on Neural Information Processing Systems}, NIPS '23,  article no. 2688. Curran Associates Inc., 2023.

\bibitem{wonderflow}
Y.~Wang, L.~Shen, Z.~You, X.~Shu, B.~Lee, J.~Thompson, H.~Zhang, and D.~Zhang.
\newblock {WonderFlow: Narration-Centric Design of Animated Data Videos}.
\newblock {\em IEEE Trans. Vis. Comput. Graph.}, pp. 1--17, 2024.

\bibitem{wang2024genartist}
Z.~Wang, A.~Li, Z.~Li, and X.~Liu.
\newblock Genartist: Multimodal {LLM} as an agent for unified image generation and editing.
\newblock In {\em The Thirty-eighth Annual Conference on Neural Information Processing Systems}, 2024.

\bibitem{jason2022cot}
J.~Wei, X.~Wang, D.~Schuurmans, M.~Bosma, B.~Ichter, F.~Xia, E.~H. Chi, Q.~V. Le, and D.~Zhou.
\newblock Chain-of-thought prompting elicits reasoning in large language models.
\newblock In {\em Proceedings of the 36th International Conference on Neural Information Processing Systems}, NIPS '22. Curran Associates Inc., 2022.

\bibitem{willett2017embedded}
W.~Willett, Y.~Jansen, and P.~Dragicevic.
\newblock Embedded data representations.
\newblock {\em IEEE Trans. Vis. Comput. Graph.}, 23(1):461--470, 2017.

\bibitem{wu2023viz2viz}
J.~Wu, J.~J.~Y. Chung, and E.~Adar.
\newblock viz2viz: Prompt-driven stylized visualization generation using a diffusion model.
\newblock {\em arXiv preprint arXiv:2304.01919}, 2023.

\bibitem{ChartInsights}
Y.~Wu, L.~Yan, L.~Shen, Y.~Wang, N.~Tang, and Y.~Luo.
\newblock {ChartInsights: Evaluating Multimodal Large Language Models for Low-Level Chart Question Answering}.
\newblock In {\em Findings of the Association for Computational Linguistics: EMNLP 2024}, pp. 12174--12200. ACL, 2024.

\bibitem{xiao2024chart}
S.~Xiao, S.~Huang, Y.~Lin, Y.~Ye, and W.~Zeng.
\newblock Let the chart spark: Embedding semantic context into chart with text-to-image generative model.
\newblock {\em IEEE Trans. Vis. Comput. Graph.}, 30(1):284–294, 2024.

\bibitem{ren2012hed}
R.~Xiaofeng and L.~Bo.
\newblock Discriminatively trained sparse code gradients for contour detection.
\newblock In {\em Advances in Neural Information Processing Systems}, vol.~25. Curran Associates, Inc., 2012.

\bibitem{xinpianchang2024yellowriver}
Xinpianchang.
\newblock What have we gone through to control the yellow river, 2024.
\newblock \url{https://www.xinpianchang.com/a12453678?kw=Guangxi\%20Nationalities\%20Museum}. Accessed: 2025-03-23.

\bibitem{yang2022designspace}
L.~Yang, X.~Xu, X.~Lan, Z.~Liu, S.~Guo, Y.~Shi, H.~Qu, and N.~Cao.
\newblock A design space for applying the freytag's pyramid structure to data stories.
\newblock {\em IEEE Trans. Vis. Comput. Graph.}, 28(1):922--932, 2022.

\bibitem{yang2023mmreact}
Z.~Yang, L.~Li, J.~Wang, K.~Lin, E.~Azarnasab, F.~Ahmed, Z.~Liu, C.~Liu, M.~Zeng, and L.~Wang.
\newblock Mm-react: Prompting chatgpt for multimodal reasoning and action.
\newblock {\em arXiv preprint arXiv:2303.11381}, 2023.

\bibitem{yao2025user}
L.~Yao, F.~Bucchieri, V.~McArthur, A.~Bezerianos, and P.~Isenberg.
\newblock User experience of visualizations in motion: A case study and design considerations.
\newblock {\em IEEE Trans. Vis. Comput. Graph.}, 31(1):174–184, 2025.

\bibitem{yao2024motion}
L.~Yao, R.~Vuillemot, A.~Bezerianos, and P.~Isenberg.
\newblock Designing for visualization in motion: Embedding visualizations in swimming videos.
\newblock {\em IEEE Trans. Vis. Comput. Graph.}, 30(3):1821--1836, 2024.

\bibitem{ye2024visatlas}
Y.~Ye, R.~Huang, and W.~Zeng.
\newblock Visatlas: An image-based exploration and query system for large visualization collections via neural image embedding.
\newblock {\em IEEE Trans. Vis. Comput. Graph.}, 30(7):3224--3240, 2024.

\bibitem{ying2022glyphcreator}
L.~Ying, T.~Tang, Y.~Luo, L.~Shen, X.~Xie, L.~Yu, and Y.~Wu.
\newblock Glyphcreator: Towards example-based automatic generation of circular glyphs.
\newblock {\em IEEE Trans. Vis. Comput. Graph.}, 28(1):400--410, 2022.

\bibitem{yu2023inpaint}
T.~Yu, R.~Feng, R.~Feng, J.~Liu, X.~Jin, W.~Zeng, and Z.~Chen.
\newblock Inpaint anything: Segment anything meets image inpainting.
\newblock {\em arXiv preprint arXiv:2304.06790}, 2023.

\bibitem{yuan2022infocolorizer}
L.-P. Yuan, Z.~Zhou, J.~Zhao, Y.~Guo, F.~Du, and H.~Qu.
\newblock Infocolorizer: Interactive recommendation of color palettes for infographics.
\newblock {\em IEEE Trans. Vis. Comput. Graph.}, 28(12):4252--4266, 2022.

\bibitem{zeng2025advancing}
X.~Zeng, H.~Lin, Y.~Ye, and W.~Zeng.
\newblock Advancing multimodal large language models in chart question answering with visualization-referenced instruction tuning.
\newblock {\em IEEE Trans. Vis. Comput. Graph.}, 31(1):525--535, 2025.

\bibitem{zhang2020dataquilt}
J.~E. Zhang, N.~Sultanum, A.~Bezerianos, and F.~Chevalier.
\newblock Dataquilt: Extracting visual elements from images to craft pictorial visualizations.
\newblock In {\em Proceedings of the 2020 CHI Conference on Human Factors in Computing Systems}, CHI '20, p. 1–13. ACM, 2020.

\bibitem{zhang2023controlnet}
L.~Zhang, A.~Rao, and M.~Agrawala.
\newblock Adding conditional control to text-to-image diffusion models.
\newblock In {\em 2023 IEEE/CVF International Conference on Computer Vision (ICCV)}, vol. 3836-3847, pp. 3813--3824, 2023.

\bibitem{zhou2024understanding}
J.~Zhou, R.~Li, J.~Tang, T.~Tang, H.~Li, W.~Cui, and Y.~Wu.
\newblock Understanding nonlinear collaboration between human and ai agents: A co-design framework for creative design.
\newblock In {\em Proceedings of the 2024 CHI Conference on Human Factors in Computing Systems}, CHI '24. ACM, 2024.

\bibitem{zhou2024stylefactory}
M.~Zhou, D.~Zhang, W.~You, Z.~Yu, Y.~Wu, C.~Pan, H.~Liu, T.~Lao, and P.~Chen.
\newblock Stylefactory: Towards better style alignment in image creation through style-strength-based control and evaluation.
\newblock In {\em Proceedings of the 37th Annual ACM Symposium on User Interface Software and Technology}, UIST '24. ACM, 2024.

\bibitem{zhou2024pictorial}
T.~Zhou, G.~Y.-Y. Chan, S.~Guo, J.~Hoffswell, C.~Xiao, V.~S.~Bursztyn, and E.~Koh.
\newblock Data pictorial: Deconstructing raster images for data-aware animated vector posters.
\newblock In {\em Proceedings of the 37th Annual ACM Symposium on User Interface Software and Technology}, UIST'24. ACM, 2024.

\bibitem{zhutian2020timeline}
C.~Zhu-Tian, Y.~Wang, Q.~Wang, Y.~Wang, and H.~Qu.
\newblock Towards automated infographic design: Deep learning-based auto-extraction of extensible timeline.
\newblock {\em IEEE Trans. Vis. Comput. Graph.}, 26(1):917--926, 2020.

\bibitem{zhutian2023sporthesia}
C.~Zhu-Tian, Q.~Yang, X.~Xie, J.~Beyer, H.~Xia, Y.~Wu, and H.~Pfister.
\newblock Sporthesia: Augmenting sports videos using natural language.
\newblock {\em IEEE Trans. Vis. Comput. Graph.}, 29(1):918--928, 2023.

\bibitem{zhutian2022viscommentator}
C.~Zhu-Tian, S.~Ye, X.~Chu, H.~Xia, H.~Zhang, H.~Qu, and Y.~Wu.
\newblock Augmenting sports videos with viscommentator.
\newblock {\em IEEE Trans. Vis. Comput. Graph.}, 28(1):824--834, 2022.

\end{thebibliography}

\end{document}